\documentclass[twocolumn]{aastex701}

\usepackage{subcaption}

\begin{document}

\newcommand{\cdunits}{cm$^{-3}$}
\newcommand{\cunits}{cm$^{-2}$}
\newcommand{\funits}{erg~cm$^{-2}$~s$^{-1}$}
\newcommand{\flunits}{erg~cm$^{-2}$~s$^{-1}$~\AA$^{-1}$}
\newcommand{\hunits}{$\rm km \ s^{-1} Mpc^{-1}$}
\newcommand{\kmps}{km s$^{-1}$}
\newcommand{\lsun}{$L_{\odot}$}
\newcommand{\lunits}{erg~s$^{-1}$}
\newcommand{\lnuunits}{erg~s$^{-1}$~Hz$^{-1}$}
\newcommand{\mbari}{$M_{\rm baryon, IGM}$}
\newcommand{\mbh}{$M_{\bullet}$}
\newcommand{\mgas}{$M_{\rm gas}$}
\newcommand{\mstar}{$M_{\ast}$}
\newcommand{\mdust}{$M_{\rm dust}$}
\newcommand{\msun}{$M_{\odot}$}
\newcommand{\msuny}{\msun yr$^{-1}$}
\newcommand{\ps}{s$^{-1}$}
\newcommand{\runits}{cts s$^{-1}$}
\newcommand{\tmone}{$^{-1}$}
\newcommand{\tmtwo}{$^{-2}$}

\newcommand{\ace}{{\it AE}}
\newcommand{\aeff}{$A_{\rm eff}$}
\newcommand{\aeffc}{$A_{\rm eff,r}$}
\newcommand{\airac}{$\alpha_{\rm IRAC}$}
\newcommand{\aet}{$\alpha_{8,24}$}
\newcommand{\alam}{$A_{\lambda}$}
\newcommand{\anuv}{$A_{\rm NUV}$}
\newcommand{\afuv}{$A_{\rm FUV}$}
\newcommand{\aox}{$\alpha_{\rm OX}$}
\newcommand{\auv}{$\alpha_{\rm UV}$}
\newcommand{\ax}{$\alpha_{\rm X}$}
\newcommand{\cs}{$\chi^2$}
\newcommand{\csn}{$\chi^2_\nu$}
\newcommand{\dflux}{$\Delta f_{\lambda}(uvm2 - NUV) / f_{\lambda} (NUV))$}
\newcommand{\fsev}{$f_{70\mu{\rm m}m}$}
\newcommand{\ftf}{$f_{24\mu{\rm m}}$}
\newcommand{\fhu}{$f_{160\mu{\rm m}}$}
\newcommand{\fx}{$f_X$}
\newcommand{\iruv}{$\nu L_{\nu,\rm 24\mu{\rm m}} /\nu L_{\nu,\it uvw2}$}
\newcommand{\lam}{$\lambda$}
\newcommand{\lamc}{$\lambda_{\rm c}$}
\newcommand{\lameff}{$\lambda_{\rm eff}$}
\newcommand{\lamlam}{$\lambda \lambda$}
\newcommand{\lamr}{$\lambda_{\rm r}$}
\newcommand{\lbol}{$L_{\rm bol}$}
\newcommand{\ledd}{$L_{\rm Edd}$}
\newcommand{\lha}{$L_{{\rm H}\alpha}$}
\newcommand{\lir}{$L_{\rm IR}$}
\newcommand{\lnu}{$L_{\nu}$}
\newcommand{\lnuks}{$L_{\nu, K_s}$}
\newcommand{\lqq}{\lq\lq}
\newcommand{\ltf}{$L_{24\mu{\rm m}}$}
\newcommand{\ltir}{$L_{\rm TIR}$}
\newcommand{\luv}{$L_{\rm UV}$}
\newcommand{\lnuv}{$L_{\nu,\rm UV}$}
\newcommand{\lnuwtwo}{$L_{\nu,\it uvw2}$}
\newcommand{\lnutf}{$L_{\nu,\rm 24\mu{\rm m}}$}
\newcommand{\logn}{$\log N$}
\newcommand{\lognt}{$\log N_{\rm thresh}$}
\newcommand{\lrest}{$\lambda_{\rm rest}$}
\newcommand{\lx}{$L_X$}
\newcommand{\lxo}{$\log (f_X/f_O)$}
\newcommand{\mhi}{$M_{\rm H I}$}
\newcommand{\nh}{$N_{\rm H}$} 
\newcommand{\nha}{$N_{{\rm H}\alpha}$} 
\newcommand{\nhgal}{$N_{\rm H}^{\rm gal}$} 
\newcommand{\nhint}{$N_{\rm H}^{\rm int}$} 
\newcommand{\rqq}{\rq\rq}
\newcommand{\sfrir}{SFR$_{\rm IR}$}
\newcommand{\sfrtot}{SFR$_{\rm TOTAL}$}
\newcommand{\sfrtf}{SFR$_{24\mu{\rm m}}$}
\newcommand{\sfruv}{SFR$_{\rm UV}$}
\newcommand{\sfrwtwo}{SFR$_{uvw2}$}
\newcommand{\sfruvir}{SFR$_{\rm UV+IR}$}

\newcommand{\todo}[1]{\Large{$\bullet$}\normalsize{(\textcolor{red}{#1})}\Large{$\bullet$}\normalsize{}}
\newcommand{\badtext}[1]{$\rightarrow$(\textcolor{red}{rewrite}) {#1} (\textcolor{red}{rewrite})$\leftarrow$}
\newcommand{\rar}{$\rightarrow$}
\newcommand{\tin}{$T_{\rm in}$}
\newcommand{\tdust}{$T_{\rm dust}$}
\newcommand{\x}{X-ray}
\newcommand{\xx}{X-rays}
\newcommand{\zabs}{$z_{\rm abs}$}
\newcommand{\zem}{$z_{\rm em}$}
\newcommand{\zmed}{$z_{\rm med}$}
\newcommand{\zla}{$z_{{\rm Ly-}\alpha}$}
\newcommand{\tocite}[1]{($\bullet$\textcolor{green}{#1}$\bullet$)}

\newcommand{\daa}{$\Delta\alpha/\alpha$}
\newcommand{\dgg}{$\Delta$G/G}
\newcommand{\dmm}{$\Delta\mu/\mu$}

\newcommand{\bt}{$b_{\rm th}$}
\newcommand{\bb}{$b_{\rm bulk}$}
\newcommand{\cloudy}{\sc cloudy}
\newcommand{\eff}{$\epsilon^{\rm ff}$}
\newcommand{\effn}{$\epsilon^{\rm ff}_{\nu}$}
\newcommand{\qal}{QAL}

\newcommand{\altwo}{Al {\sc \,ii}}
\newcommand{\althree}{Al {\sc \,iii}
\newcommand{\catwo}{Ca{\sc \,ii}}
\newcommand{\cfour}{C{\sc \,iv}}
\newcommand{\ctwo}{C{\sc \,ii}}
\newcommand{\ctwof}{$[$C\sc{ \,ii}$]$}
\newcommand{\cthree}{C{\sc \,iii}}
\newcommand{\cones}{C{\sc \,i}$^{\ast}$}
\newcommand{\ctwos}{C{\sc \,ii}$^{\ast}$} }
\newcommand{\feka}{Fe K$\alpha$}
\newcommand{\fetwo}{Fe{\sc \,ii}}
\newcommand{\ha}{H$\alpha$}
\newcommand{\hb}{H$\beta$}
\newcommand{\heone}{He{\sc \,i}}
\newcommand{\hetwo}{He{\sc \,ii}}
\newcommand{\hone}{H{\sc \,i}}
\newcommand{\htwo}{H{\sc \,ii}}
\newcommand{\mhtwo}{H$_2$}
\newcommand{\mgone}{Mg{\sc \,i}}
\newcommand{\mgtwo}{Mg{\sc \,ii}}
\newcommand{\mntwo}{Mn{\sc \,ii}}
\newcommand{\naone}{Na{\sc \,i} }
\newcommand{\nfive}{N{\sc \,v}}
\newcommand{\natwo}{Na{\sc \,ii} }
\newcommand{\ntwo}{N{\sc \,ii}}
\newcommand{\ntwof}{[N{\sc \,ii}]}
\newcommand{\oone}{O{\sc \,i}}
\newcommand{\osix}{O{\sc \,vi}}
\newcommand{\othree}{O{\sc \,iii}}
\newcommand{\othreef}{[O{\sc \,iii}]}
\newcommand{\otwof}{[O{\sc \,ii}]}
\newcommand{\oonef}{[O{\sc \,i}]}
\newcommand{\sifour}{Si{\sc \,iv}}
\newcommand{\sifive}{Si{\sc \,v}}
\newcommand{\sithree}{Si{\sc \,iii}}
\newcommand{\sitwo}{Si{\sc \,ii}}
\newcommand{\stwo}{S{\sc \,ii}}
\newcommand{\stwof}{[S{\sc \,ii}]}
\newcommand{\titwo}{Ti{\sc \,ii}}

\newcommand{\lya}{Ly-$\alpha$}
\newcommand{\lyb}{Ly-$\beta$}
\newcommand{\lyaf}{Ly-$\alpha$ forest}

\newcommand{\chandra}{\textit{Chandra}}
\newcommand{\fuv}{{\it FUV}}
\newcommand{\galex}{{GALEX}}
\newcommand{\hr}{{HIRES}}
\newcommand{\herschel}{\textit{Herschel}}
\newcommand{\hri}{{HRI}}
\newcommand{\hst}{{\it HST}}
\newcommand{\iras}{{\it IRAS}}
\newcommand{\iso}{{\it ISO}}
\newcommand{\nuv}{{\it NUV}}
\newcommand{\rosat}{{\it ROSAT}}
\newcommand{\spitzer}{\textit{Spitzer}}
\newcommand{\swift}{\textit{Swift}}
\newcommand{\twodf}{{\it 2dfGRS}}
\newcommand{\xmm}{\textit{XMM-Newton}}
\newcommand{\wone}{{\it uvw1}}
\newcommand{\wtwo}{{\it uvw2}}
\newcommand{\wtwoc}{{\it uvw2$\,^{\prime}$}}
\newcommand{\mtwo}{{\it uvm2}}
\newcommand{\uvot}{{UVOT}}
\newcommand{\wise}{{\it WISE}}
\newcommand{\rxte}{{\it RXTE}}
\newcommand{\nustar}{\textit{NuSTAR}}
\newcommand{\asca}{{\it ASCA}}
\newcommand{\einstein}{{\it EINSTEIN}}
\newcommand{\ginga}{{\it GINGA}}
\newcommand{\bbxrt}{{\it BBXRT}}
\newcommand{\suzaku}{{\it Suzaku}}
\newcommand{\sax}{{\it BeppoSAX}}
\newcommand{\fermi}{{\it Fermi}}
\newcommand{\integral}{\textit{INTEGRAL}}
\newcommand{\astroh}{{\it ASTRO-H}}
\newcommand{\heao}{{\it HEAO 1}}
\newcommand{\hitomi}{\textit{Hitomi}}
\newcommand{\xarm}{\textit{XRISM}}
\newcommand{\xrism}{\textit{XRISM}}
\newcommand{\athena}{\textit{Athena}}
\newcommand{\lynx}{\textit{Lynx}}
\newcommand{\planck}{\textit{Planck}}

\newcommand{\hl}{\hspace{-0.7cm}}

\newcommand{\idl}{{\sc idl}}
\newcommand{\iraf}{{\sc iraf}}
\newcommand{\surphot}{{\sc surphot}}
\newcommand{\vp}{{\sc vpfit}}

\newcommand{\er}{Equation~\ref}
\newcommand{\fr}{Fig.~\ref}
\newcommand{\scr}{Sec.~\ref}
\newcommand{\tr}{Table~\ref}

\newcommand{\exi}{\begin{equation}}
\newcommand{\exo}{\end{equation}}

\newcommand{\aer}[3]{$#1^{#2}_{#3}$} 
\newcommand{\scier}[4]{$#1^{#3}_{#4} \times 10^{#2}$} 
\newcommand{\ten}[2]{$#1\times 10^{#2}$} 

\newcommand{\nB}{\noindent$\bullet$}
\newcommand{\todon}[1]{{\Large $\bullet$}{\normalsize \bf{#1}}}
\newcommand{\bp}{{\bf(p)}}
\newcommand{\bs}{{\bf(s)}}

\newcommand{\mro}{\multirow}
\newcommand{\mrr}[1]{\multirow{2}{*}{#1}}
\newcommand{\mrrr}[1]{\multirow{3}{*}{#1}}
\newcommand{\mrf}[1]{\multirow{5}{*}{#1}}

\newcommand{\cgr}{\color{grey}}
\newcommand{\cred}{\color{red}}
\newcommand{\bcred}[1]{\color{red}{\bf #1}}
\newcommand{\bcblue}[1]{\color{blue}{\bf #1}}



\title{Cross-Calibration of Galaxy Cluster Temperatures Measured with \nustar, \xmm, and \chandra}

\author[0009-0005-5605-4686]{Fiona Lopez}
\email{fionalopez687@gmail.com}
\affiliation{Department of Physics \&Astronomy, The University of Texas A\&M, 578 University Dr, College Station, TX 77843, USA}

\author[0000-0001-9110-2245]{Daniel R. Wik}
\email{wik@astro.utah.edu}
\affiliation{Department of Physics \& Astronomy, The University of Utah, 115 South 1400 East, Salt Lake City, UT 84112, USA}

\author[0009-0003-7738-9173]{Cicely Potter}
\email{potter.cicely@gmail.com}
\affiliation{Department of Physics \& Astronomy, The University of Utah, 115 South 1400 East, Salt Lake City, UT 84112, USA}

\author[0000-0002-8882-6426]{Randall A. Rojas Bolivar}
\email{rrojasbo94@gmail.com}
\affiliation{Department of Physics \& Astronomy, The University of Utah, 115 South 1400 East, Salt Lake City, UT 84112, USA}

\author[0000-0002-3132-8776]{Ayşegül Tümer}
\email{aysegultumer@gmail.com}
\affiliation{Center for Space Science and Technology, University of Maryland, Baltimore County (UMBC), Baltimore, MD 21250, USA}
\affiliation{NASA Goddard Space Flight Center, Greenbelt, MD 20771, USA}  
\affiliation{Center for Research and Exploration in Space Science and Technology (CRESST II), Greenbelt, MD 20771, USA}

\author[0000-0001-7917-3892]{Dominique Eckert}
\email{Dominique.Eckert@unige.ch}
\affiliation{Department of Astronomy, University of Geneva, ch. d’Écogia 16, CH-1290 Versoix Switzerland}

\author[0000-0002-9112-0184]{Fabio Gastaldello}
\email{fabio.gastaldello@inaf.it}
\affiliation{INAF – IASF Milano, via A. Corti 12, I-20133 Milano, Italy}

\author[0000-0002-1984-2932
]{Brian W Grefenstette}
\email{bwgref@srl.caltech.edu}
\affiliation{Cahill Center for Astrophysics, California Institute of Technology, 1216 East California Boulevard, Pasadena, CA 91125, USA}

\author[0000-0003-1252-4891]{Kristin Madsen}
\email{kristin.c.madsen@nasa.gov}
\affiliation{X-Ray Astrophysics Laboratory, NASA Goddard Space Flight Center, Greenbelt, MD 20771, USA}

\author[
0000-0003-0791-9098]{Ben Maughan}
\email{ben.maughan@bristol.ac.uk}
\affiliation{H.H. Wills Physics Laboratory, University of Bristol, Tyndall Avenue, Bristol, BS8 1TL, UK}

\author[
0000-0002-3031-2326
]{Eric D. Miller}
\email{milleric@mit.edu}
\affiliation{Kavli Institute for Astrophysics and Space Research, Massachusetts Institute of Technology, 77 Massachusetts Avenue, Cambridge, MA 02139, USA}

\author[0000-0002-4962-0740]{Gerrit Schellenberger}
\email{gerrit.schellenberger@cfa.harvard.edu}
\affiliation{Center for Astrophysics, Harvard \& Smithsonian, 60 Garden Street, Cambridge, MA 02138, USA}

\author{A. N. Wallbank}
\email{aw16763.2016@my.bristol.ac.uk}
\affiliation{1H. H. Wills Physics Laboratory, University of Bristol, Tyndall Ave, Bristol BS8 1TL, UK}



\begin{abstract}
The use of galaxy clusters to constrain cosmology is limited in part due to uncertainties in derived cluster masses, which often 
depend on the gas temperature.
Unfortunately, 
there exists a longstanding discrepancy in temperature measurements of the same galaxy clusters made by the two most sensitive X-ray observatories, \chandra\ and \xmm.
The \nustar\ X-ray Observatory's greater sensitivity to the exponential turnover in the bremsstrahlung continuum allows for more precise and potentially more accurate galaxy cluster temperature estimates, especially given its unique ability to independently calibrate its optics in orbit. 
We present new \nustar\ spectra of 10 relaxed ($5~{\rm keV} < kT < 10$~keV) clusters, extracted from identical regions as previous spectra from \chandra\ and \xmm.
The 3--20~keV spectra are well fit by single temperature models, and fits done in narrower bandpasses provide no clear evidence in support of the existence of multi-temperature gas. 
We find \nustar\ temperatures are typically $\sim$15\% higher than \xmm\ temperatures.
In contrast, good agreement is found between \nustar\ and \chandra\ temperatures for clusters with $kT\lesssim 7$~keV, with \chandra\ measurements exceeding \nustar's in hotter systems.
When 
more clusters
are included, the trend is reinforced and can be extended to higher temperatures.
A generic increase to \chandra's $E>2$~keV effective area ($\sim$5\% at 5~keV)
is found to explain
the trend reasonably well.
These results demonstrate the potential for \nustar\ data to address the two-decade old temperature discrepancy between \chandra\ and \xmm.

\end{abstract}



\section{Introduction}
\label{sec:intro}

Galaxy clusters are the most massive gravitationally-bound structures in the universe. The number of clusters as a function of mass (and how that increases over time) can be used to precisely quantify dark matter and other cosmological parameters \citep[e.g.,][]{Allen_2011,Diehl_2021,Aymerich}.
The eROSITA X-ray survey has the goal of detecting approximately 100,000 galaxy clusters in order to study cosmology. There are multiple methods of estimating masses from the data in this survey \citep{ghirardini2024srgerosita}. One approach uses hydrostatic mass measurements, which can be calculated from the pressure or temperature measurements \citep[e.g.,][]{Ettori_2013}. 
Although cosmological constraints using galaxy clusters do not require detailed hydrostatically-derived masses, studies typically use scaling relations with related quantities, such as cluster temperature and luminosity, even when not directly measured
\citep[e.g.,][]{XXLcosmo,ghirardini2024srgerosita}.

When the hot plasma between galaxies of the cluster, the intracluster medium (ICM), is in hydrostatic equilibrium, cluster gravitational potential energy is thermalized \citep[e.g.,][]{Scheck_2023}.
X-ray telescopes are ideal for measuring both the temperature and density via the bremsstrahlung emission spectrum of the ICM.


Mass estimations under the hydrostatic equilibrium assumption becomes inaccurate if sources of non-thermal pressure support, such as turbulence, bulk motions, and magnetic fields, are present in the system \citep{Ettori_2021}. 
\cite{Pearce2020} explained how the non-thermal pressure is expected to contribute as much as 30\% of the overall pressure in the outskirts of galaxy clusters currently. Their results are based on numerical simulations.


Another, non-physical origin of bias for hydrostatic mass estimates is the calibration of X-ray telescopes that leads to a mass uncertainty at a level of 10\textendash20\% \citep[e.g.,][]{Wik,Nevalainen,Kettula}.
In particular, there is a long-standing systematic difference in temperature measurements of the same clusters by the two flagship X-ray observatories, \chandra\ and \xmm; specifically, we are interested in the discrepancy between the spectro-imaging instruments of each observatory: ACIS-I and ACIS-S (used without a grating) for \chandra---which generally return consistent measurements to each other---and the three EPIC instruments (MOS-1, MOS-2, and pn) of \xmm.
The three instruments of \xmm\ also exhibit some systematic differences in their temperature measurements, but their spectra are often fit simultaneously to yield a combined, more precise average temperature estimate, which is what we consider here.
These observatories provide high signal-to-noise X-ray spectra that allow for detailed and sensitive measurements of the ICM properties. However, these instruments strongly depend on the broadband calibration of a telescope's effective area. The effective area of the X-ray telescopes needs to be properly calibrated in order to prevent distortion of the slope of observed X-ray spectra which would potentially bias quantities e.g., the ICM temperatures derived from spectral fits (\cite{Potter}). If the temperatures of the ICM were systematically underestimated then this would lead to underestimates of hydrostatic masses.



International Astrophysical Consortium for High Energy Calibration (IACHEC) is a cross-calibration effort between the X-ray observatories and galaxy clusters is one of the calibration sources that are used. 
As part of the IACHEC effort, \cite{Nevalainen} and \cite{Schellenberger} focused on cross-calibration uncertainties using galaxy cluster measurements from \chandra\ and \xmm.
\cite{Migkas_2024} presented a cross-calibration between eROSITA and \chandra, \xmm\ temperature measurements. They found that eROSITA shows a strong discrepancy with \chandra, in the hard band eROSITA resulted in approximately 25--60\% lower measurements than \chandra\ for temperatures approximately 2--10 keV. 
When compared to \xmm\ in the full band, eROSITA measured 10-28\% lower than \xmm\ for temperatures approximately 2--7 keV.

\citet{Schellenberger} undertook a detailed characterization of the discrepancies between \chandra\ and \xmm.
They found that \chandra/ACIS had systematically higher temperatures compared to \xmm/EPIC,
and that the discrepancy increased as cluster temperature increased. While temperature differences are typically on the order of 10\% for moderate temperature systems,
\chandra\ temperatures were found to be $\sim$30\% higher than \xmm\ for clusters with $kT \sim 10$~keV. 
The observed temperature differences were determined not to be explained by physical effects, such as line-of-sight multi-temperature structure.
While measurements made with data only at high ($E>2$~keV) energies were more consistent, the full band temperature discrepancy appears to be driven primarily by the systematic effective area calibration differences at soft ($0.7~{\rm keV}<E<2$~keV) energies, where single temperature fits to \xmm\ spectra exhibit soft excess emission \citep{Nevalainen,Schellenberger}.

The hard X-ray observatory, \nustar\ ($3~{\rm keV}<E<79$~keV), is more sensitive to the exponential turnover in the thermal continuum, allowing more precise and potentially more accurate galaxy cluster temperature estimates for massive clusters. \nustar\ has a larger effective area (3\textendash20~keV) compared to the 
2\textendash7 keV energy band of \chandra\ and \xmm, and measurements can effectively be made in a 3\textendash20~keV bandpass.
\cite{Wallbank} performed the first cross-calibration between \nustar\ and \chandra\ temperatures in the ICM of clusters.
They found \chandra\'s temperature measurements were systematically higher than those of \nustar, precisely the opposite trend one would expect from physical or projection effects.
The average \nustar\ temperature measurement recorded was 10\textendash15\% lower than \chandra\ for a cluster temperature of $kT\sim10$~keV.
Given the sense of the discrepancy, 
inconsistent cross calibration between the 
observatories is the only viable explanation. 

The results from these studies on the cross-calibration between the X-ray observatories motivate further understanding the nature of the temperature discrepancies. In this paper, we present comparisons of galaxy cluster temperatures measured with \nustar\ and \chandra\ and \xmm.
Using short ($\sim$20~ks) \nustar\ observations of a sample of 10 galaxy clusters, we extend temperature measurements available with \nustar\ of clusters to lower overall temperatures, and compare with previous measurements of the same clusters with \chandra. While \cite{Wallbank} primarily focused on merging galaxy clusters, this study focuses mainly on relaxed galaxy clusters.

All uncertainties are quoted at the {\bf 68\%} confidence levels unless otherwise stated. 



\section{Observations} \label{sec:style}

We selected clusters for this sample using various criteria, most important of which was the existence of previous, high precision \chandra\ and \xmm\ temperature measurements. A natural source for such clusters is the flux-limited HIFLUGCS catalog \citep{Reiprich_2002}, which includes the X-ray brightest clusters in the sky and was the sample used in \citet{Schellenberger}. 
We also restricted the clusters to the German (MPE) half of the eROSITA-DE sky, since data are released.
The clusters were additionally required to be included in the Planck cluster catalog \citep{Planck2016} to further enhance the legacy value of the data.
The resulting sample of 11 clusters have temperature profiles that gradually rise from the X-ray emission center until they reach a peak, followed by a relatively flat temperature profile for several arcminutes in radius, allowing large, mostly isothermal regions to be used for measurements.
As part of the \nustar\ Cycle 8 Priority C GO program (PI Wik), all of the clusters except Abell~3562 were observed, leaving a final sample of 10 relaxed clusters following the criteria of \cite{Zhang_2011}.

Relaxed galaxy clusters are better suited for the characterization of temperature differences between observatories. Merging galaxy clusters contain generally hotter gas with shocked gas fractions that cause the temperature distribution in the gas to span a wider range. This broader range can worsen bandpass weighting effects that impact the average measured temperature by an observatory \citep{Mazzotta}. 
Merging clusters are thus a less ideal place to study temperature cross-calibrations. 
The focus on relaxed clusters can help avoid bias from outliers not due to cross-calibration effects, which would impact the relationship found between \nustar\ and \chandra\ or \xmm.
However, azimuthal temperature fluctuations are present even in relaxed clusters, potentially biasing low globally-measured temperatures by $\sim$0.5~keV \citep{Lovisari+24}.

In addition, this sample significantly expands the number of relaxed and moderate mass systems observed by \nustar, which provides opportunities for future studies.
Over half of the clusters show a moderate amount of temperature variation due to the presence of sloshing cores or other weak merger signatures, for example.
These features can be further investigated individually for future archival studies, although these regions have been excluded from the spectra considered here. 

The data for this study were taken from the \nustar\ public archive.
\nustar\ observations result in two datasets from two telescopes, designated by their focal plane modules (FPM), A and B, for each cluster. 
The \chandra\ and \xmm\ observations result from the instruments \chandra/ACIS and \xmm/EPIC-PN combined \cite{Schellenberger}, respectively.

We report the specifications of each set of observations for each cluster in Table~\ref{observ}.

\begin{deluxetable*}{lccccc}
\tabletypesize{\scriptsize}
\tablewidth{0pt}
\tablehead{
\colhead{ } & \colhead{\nustar} & \colhead{Redshift} & \colhead{RA} &
\colhead{Dec} & $t_{\rm exp}$\\
\colhead{Cluster Name} & \colhead{ObsID} & \colhead{z} & \colhead{degrees} &
\colhead{degrees} & (ks)
}
\startdata
        RXC J1504  & 70860001002  & 0.2172 & 226.032 & -2.804 & 23.7 \\
        Abell 3571 & 70860002002 &  0.0390 & 206.866 & -32.857 & 18.0 \\
        Abell 3558 & 70860003002 &  0.0484 & 201.987 & -31.495 & 20.8 \\
        Abell 1651 & 70860004002 & 0.0850 & 194.841 & -4.195 & 19.5 \\
        Abell 3391 & 70860005002 & 0.0561 & 96.595 & -53.683 & 19.7 \\
        Abell 1650 & 70860006002 & 0.0838 & 194.673 & -1.762 & 18.5 \\
        Abell 3158 & 70860007002 & 0.0592 & 55.722 & -53.627 & 22.1 \\
        Abell 3112 & 70860008002 & 0.0753 & 49.490 & -44.238 & 22.2 \\
        Abell 1644 & 70860009002 & 0.0474 & 194.299 & -17.409 & 21.1 \\
        Abell 496  & 70860010002 & 0.0331 & 68.408 & -13.261 & 19.9 \\
\enddata
\caption{\label{observ} Observations of sample of relaxed clusters. The columns denote the cluster name, \nustar\ ObsID, redshift, R.A, decl., and total cleaned exposure time. The R.A. and decl. are the centers of the extraction region used for spectral fitting. \chandra\ and \xmm\ ObsID's are available in \cite{Schellenberger}.}
\end{deluxetable*}

\section{\nustar\ Data Preparation}
\label{sec:proc}

\subsection{Light-curve Filtering and Data Processing}
\label{sec:proc:proc}

The \nustar\ observations of each cluster were processed using NUSTARDAS version 2.0.0 with the CALDB version (20220525) and Heasoft version (6.30.1). Both FPM data, A and B, were processed. 

Initial images were extracted from the cleaned event lists FPMA and FPMB data; these images were then combined to assess source extent relative to the background level. 
High background periods may be present due to solar flares and changes in the geographic location of the South Atlantic Anomaly and the Tentacle, which should be removed through light curve filtering. 
Light curves from the entire field of view, excluding cluster and any point source emission, were binned by 100~s and filtered by creating a user specified good time interval (GTI). 
High or low rate light curve bins were manually identified and then excluded from the new GTI file; this process was done separately in hard (50\textendash160~keV) and soft (1.6\textendash10~keV) bands to capture flares in both the particle background and direct solar emission, respectively.
The FPMA and FPMB GTIs were determined independently, although excluded times are generally similar.
The data were then reprocessed using the new GTI files with {\tt nupipeline}. 
Images were then remade from the newly reprocessed data, and all further data products are derived from these event files.  

\subsection{\nustar\ Background Characterization}
\label{sec:proc:bgd}

\nustar\ has a stable background that was characterized using an empirical model incorporated in {\tt nuskybgd} \citep{Wik}.
{\tt nuskybgd} fits this model to extracted background spectra from multiple regions and can then produce a simulated background for any region of interest. The main contributions to the \nustar\ background can be characterized as originating from the instrument Compton scattered continuum emission, instrument emission lines, cosmic X-ray background (CXB) from the sky leaking past the aperture stops, reflected solar x-rays, and focused/ghost CXB\footnote{\url{https://heasarc.gsfc.nasa.gov/docs/nustar/NuSTAR_observatory_guide-v1.0.pdf}}.
Background regions consisted of 6\arcmin\ square boxes centered on each of the four detectors in each focal plane with an exclusion region centered on the cluster position with a 3.5\arcmin\ radius. 
We chose a background region just outside the spectral extraction region we used for the the cross-calibration measurement (see Section~\ref{sec:proc:spec}). Since detectable cluster emission is still present in the background regions (outside of the exclusion regions), {\tt APEC} \textbf{version 3.0.9}
models were added to the default background model so that emission would not be confused with a background component (\cite{Smith_2001, Foster12}). 

\begin{figure}
    \centering
    \includegraphics[width=\linewidth]{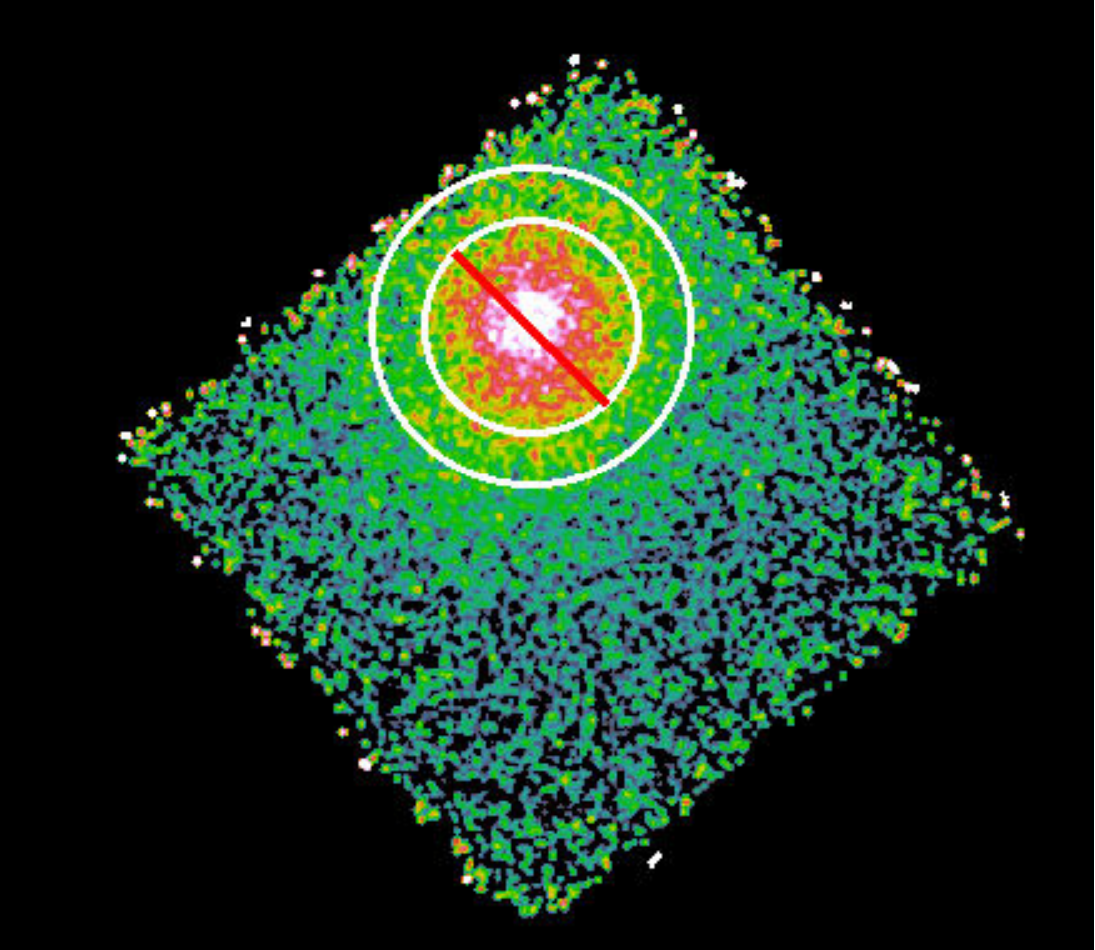}
    \caption {\nustar\ image of A496, ObsID 70860010002 showing the {\tt nuskybgd} background subtracted and exposure corrected image in the 3\textendash20 keV band. The inner regions is the exclusion region. The outer region represents where the source spectra was extracted.}
    \label{fig:abell_0496}
\end{figure}

All of the spectra were fit using {\tt Xspec} version 12.12.1. The background spectra were manually fit until reasonably best fit parameters were found.
Typically, the components due to the CXB were first fixed to the nominal expected flux, allowing the many other mostly instrumental background components to get close to their best-fit values.
The ICM temperature and abundance were also initially fixed to the global temperature from \cite{Schellenberger} and 0.3 solar, respectively, although the normalizations were left free (and independent of each other). The Anders and Grevesse solar abundances were used for all fits ({\tt Xspec} table {\tt angr}; \cite{Anders}).
The aperture CXB normalization was then thawed along with the temperature and abundance.
At this stage, the redshift was fixed to the fiducial value given in the NASA Extragalactic Database (NED) for each cluster.
The focused CXB was left fixed, as it is sufficiently faint and has a similar spectral shape as the ICM component, so any difference in brightness will be incorporated into that component.

Once the background has been characterized, the resulting model can produce background images in any bandpass and background spectra for any region. The background images were produced in the 4\textendash20 keV bandpass. Exposure maps were created through {\tt nuexpomap} at 10 keV energy. The background images and exposure maps were used to create images such as Figure \ref{fig:abell_0496}.

\begin{figure}
    \centering
    \includegraphics[width=\linewidth]{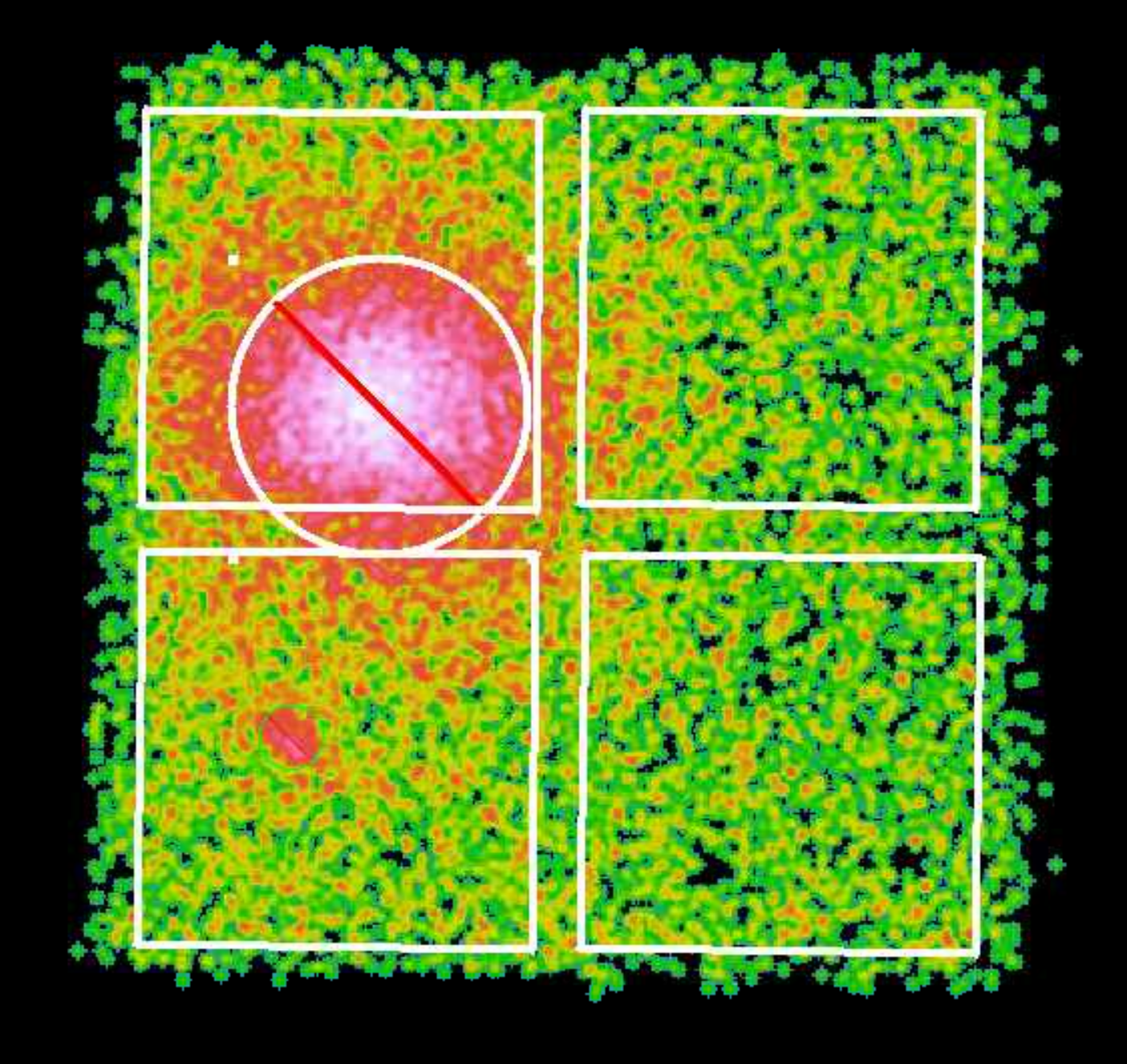}
    \caption {\nustar\ FPMA image of A1651 in the 3\textendash20 keV energy band, ObsID 70860004002 showing the exclusion region used to model the background with {\tt nuskybgd}. This image is from detector FPMA in the 3\textendash20 keV energy band.}
    \label{fig:abell3571}
\end{figure}

\subsection{Global Spectral Extraction and Fitting}
\label{sec:proc:spec}

Global regions were created with a radius of 3.5\arcmin\ around the core of the cluster (for examples, see~Fig.~\ref{fig:abell3571} \& \ref{fig:abell_0496}) for straightforward comparison with the temperature measurements of \cite{Schellenberger}. 
For the cool core clusters in the sample, an inner region was excluded from the region; see Section~\ref{sec:proc:psfscat} for details. 
The global spectra and response files appropriate for an extended source were then generated through the task {\tt nuproducts}.
Corresponding background spectra were generated from the model, described in the Section~\ref{sec:proc:bgd}, using {\tt nuskybgd}.
These steps were separately done for both the FPMA and FPMB datasets.
For each cluster, both A and B spectra were simultaneously fit with {\tt Xspec} by minimizing the C-stat statistic (command {\tt statistic cstat}, but since background is being subtracted, {\tt Xspec} technically uses the modified w statistic). The normalization, temperature, and abundance were free to vary when fitting, while the redshift was fixed to the same fiducial value, used in the background characterization step, provided by NED for each cluster. 
All parameters for the FPMA and FPMB detectors were tied to each other when fitting.
A gain shift was also fit as part of the modeling.
Allowing the gain to vary provides for any mismatch between the accepted redshift of these well-studied clusters (see Table \ref{observ}) and the location of the iron complex without adversely affecting the temperature estimate, since an artificial redshift will apply a multiplicative shift instead of a uniform shift more likely due to a slight gain offset
\citep{Rojas_2023}. 
Gain offsets of this type have also been seen in observations of AGN, may result from delayed low energy calibration over time, and are under active investigation (Grefenstette, private communication
For example, the Cycle 4 observation of Abell~2146 originally showed a large gain offset of $<-0.1$~keV, but when recently reanalyzed with current software and CALDB, the required shift shrinks by more than a factor of 2.
The application of a gain offset recovers consistent temperatures regardless of the state of the calibration, but in any case the impact on the measured temperature is on the order of the statistical uncertainty.


\begin{figure*}
    \begin{subfigure}[t]{0.5\linewidth}
        \includegraphics[width=0.72\linewidth, angle =270]{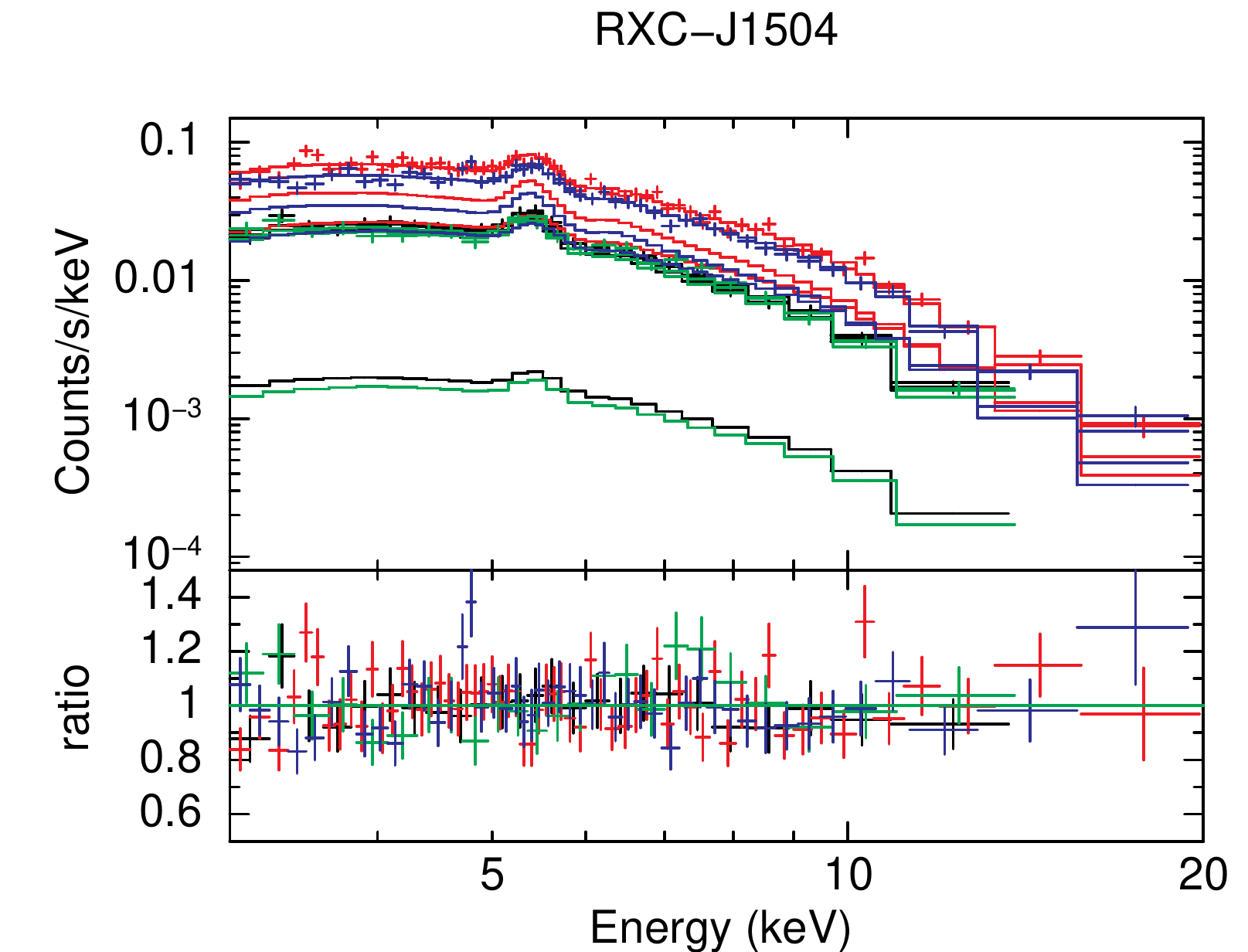} 
    \end{subfigure}
    \hfill
    \begin{subfigure}[t]{0.5\linewidth}
        \includegraphics[width=0.72\linewidth, angle =270]{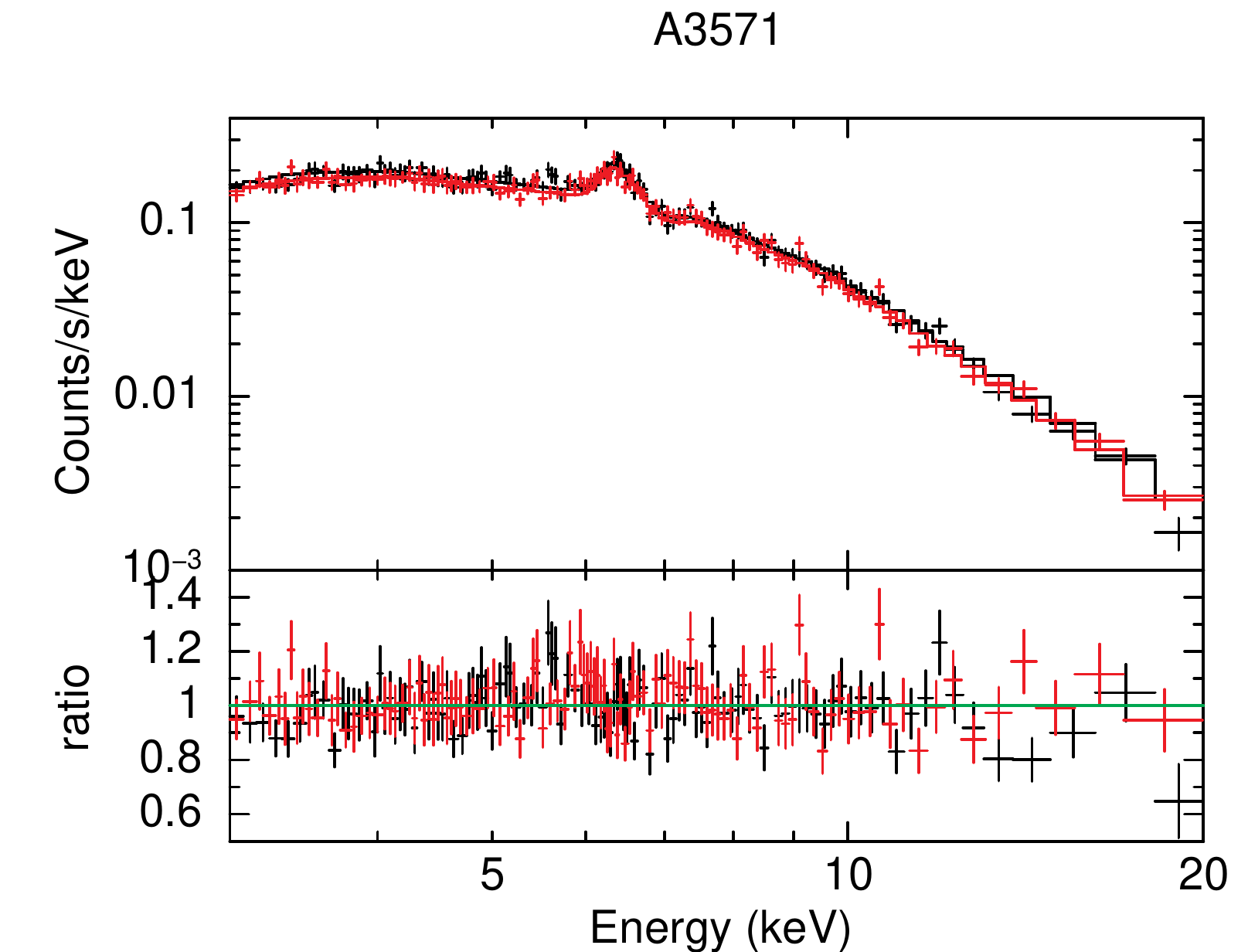} 
    \end{subfigure}
    \par\vspace{0.5cm}\par
    \begin{subfigure}[t]{0.5\linewidth}
        \centering
        \includegraphics[width=0.72\linewidth, angle =270]{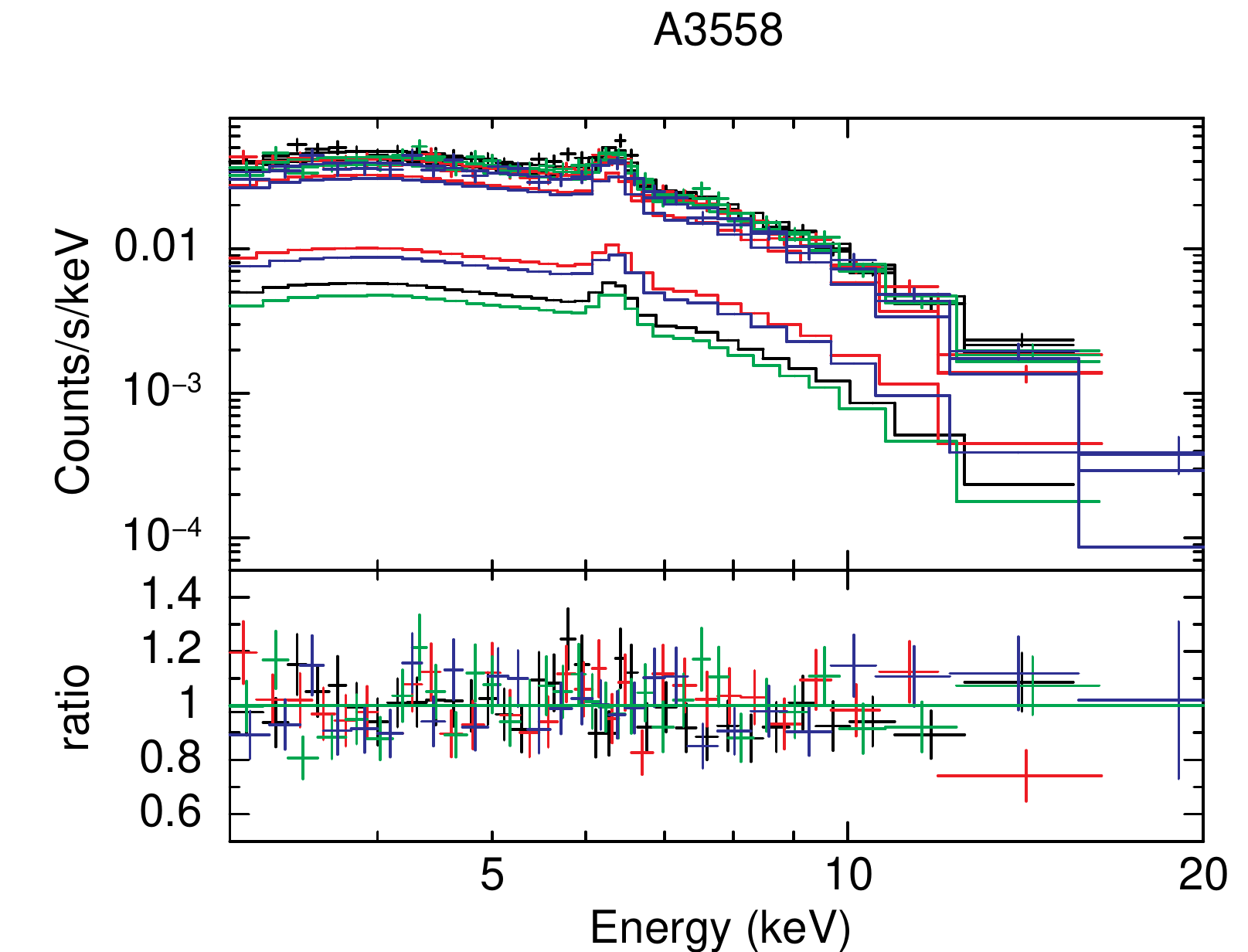} 
    \end{subfigure}
    \hfill
    \begin{subfigure}[t]{0.5\linewidth}
        \centering
        \includegraphics[width=0.72\linewidth, angle =270]{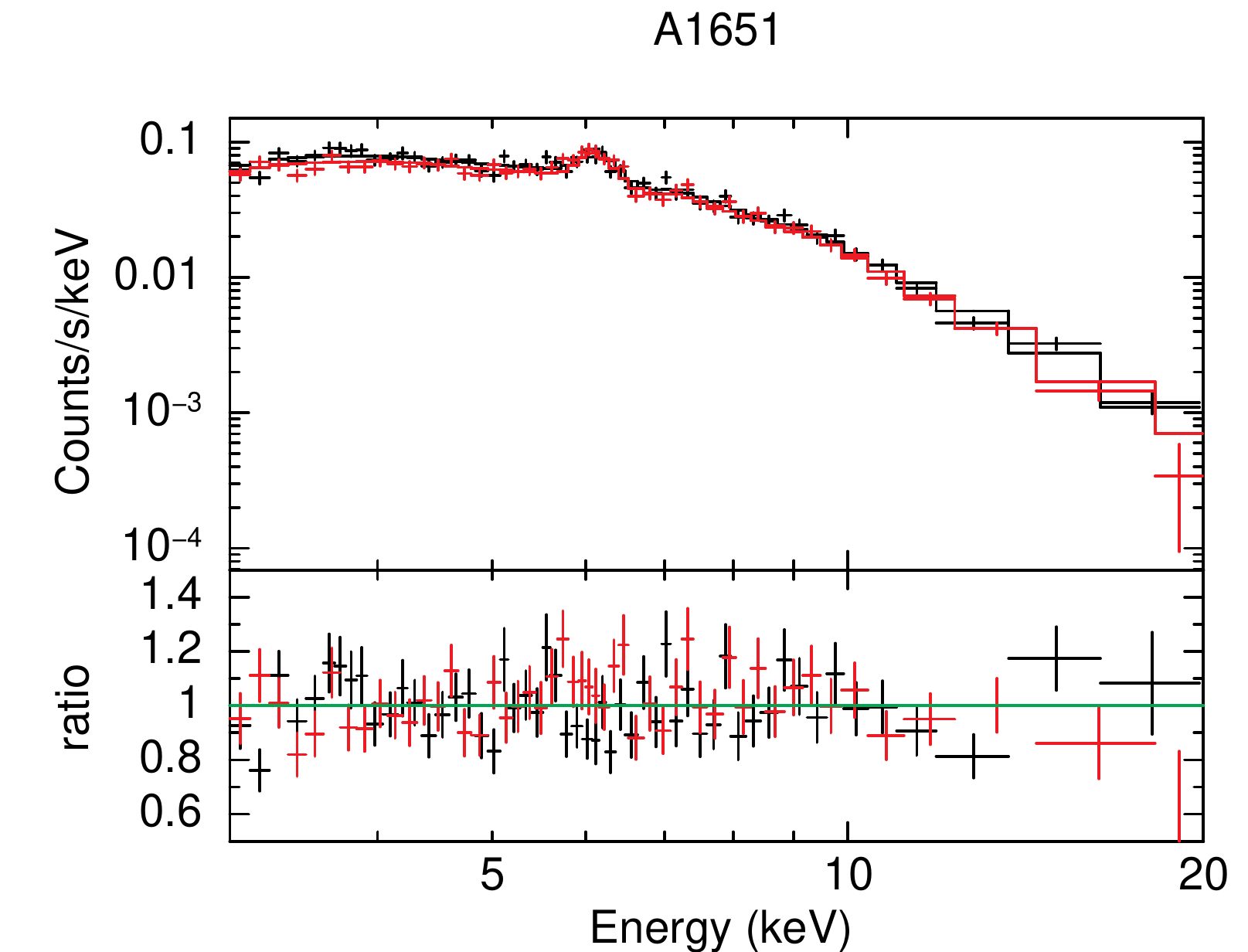} 
    \end{subfigure}
    \par\vspace{0.5cm}\par
    \begin{subfigure}[t]{0.5\linewidth}
        \centering
        \includegraphics[width=0.72\linewidth, angle =270]{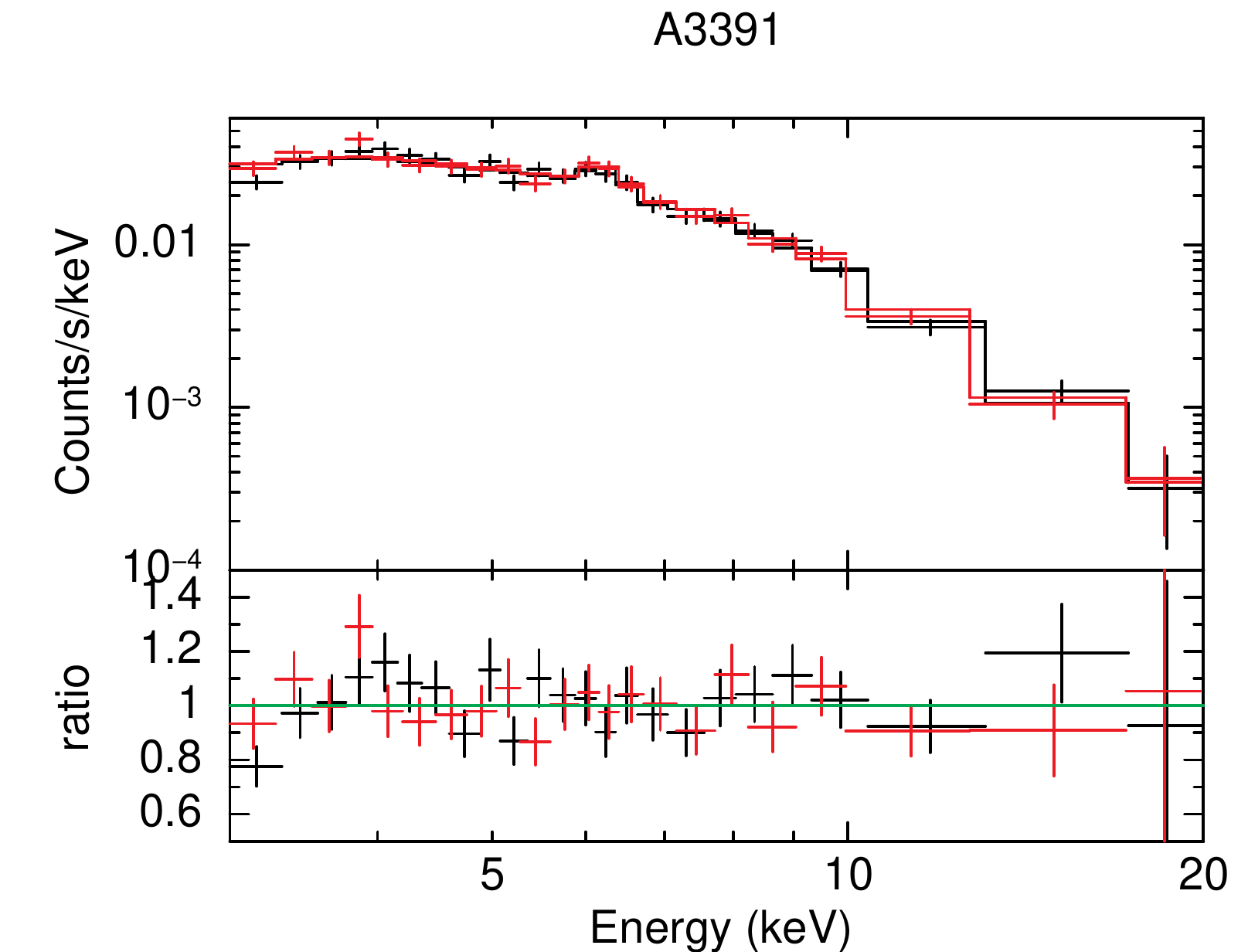} 
    \end{subfigure}
    \hfill
    \begin{subfigure}[t]{0.5\linewidth}
        \centering
        \includegraphics[width=0.72\linewidth, angle =270]{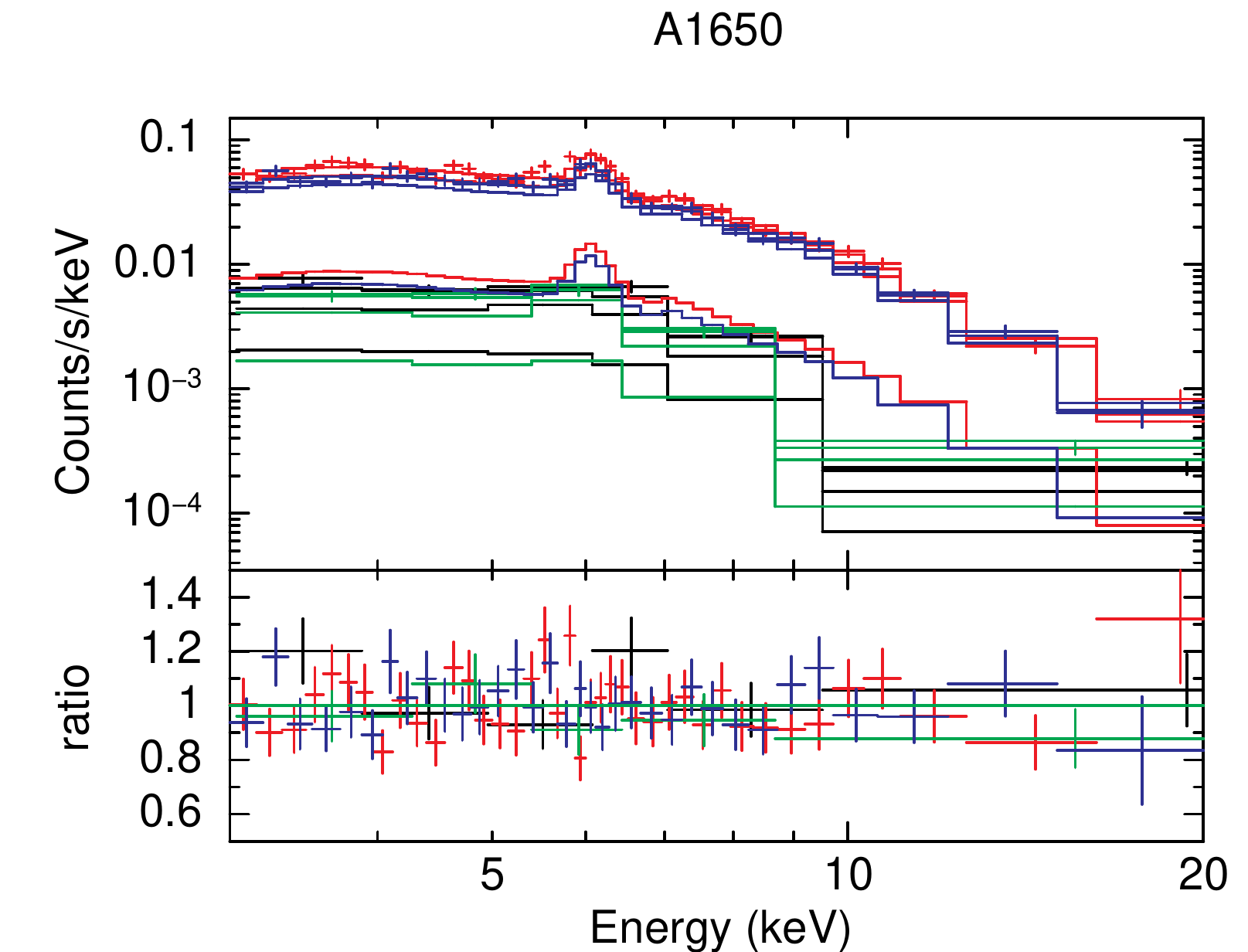} 
    \end{subfigure}
    \caption{\nustar\ spectral fits of clusters RXC~J1504 (top left), A3571 (top right), A3558 (middle left), A1651 (middle right), A3391 (bottom left), and A1650 (bottom right).
    FPMA and FPMB spectra and their corresponding models are drawn in black and red, respectively.
    For the cool core clusters, the black/green spectra correspond to the outer annulus, and the red/blue spectra to the inner annulus---  corresponding to FPMA/FPMB---and there are two sets of models, with corresponding colors according to spectrum, that describe the source emission both from that region and that scattered into it from the other region, along with the sum of those models passing through the data.
    The bottom panels show the ratio of the data to the total model. }
    
    
    \label{fig:spec1}
\end{figure*}

\begin{figure*}
    \begin{subfigure}[t]{0.5\linewidth}
        \includegraphics[width=0.72\linewidth, angle =270]{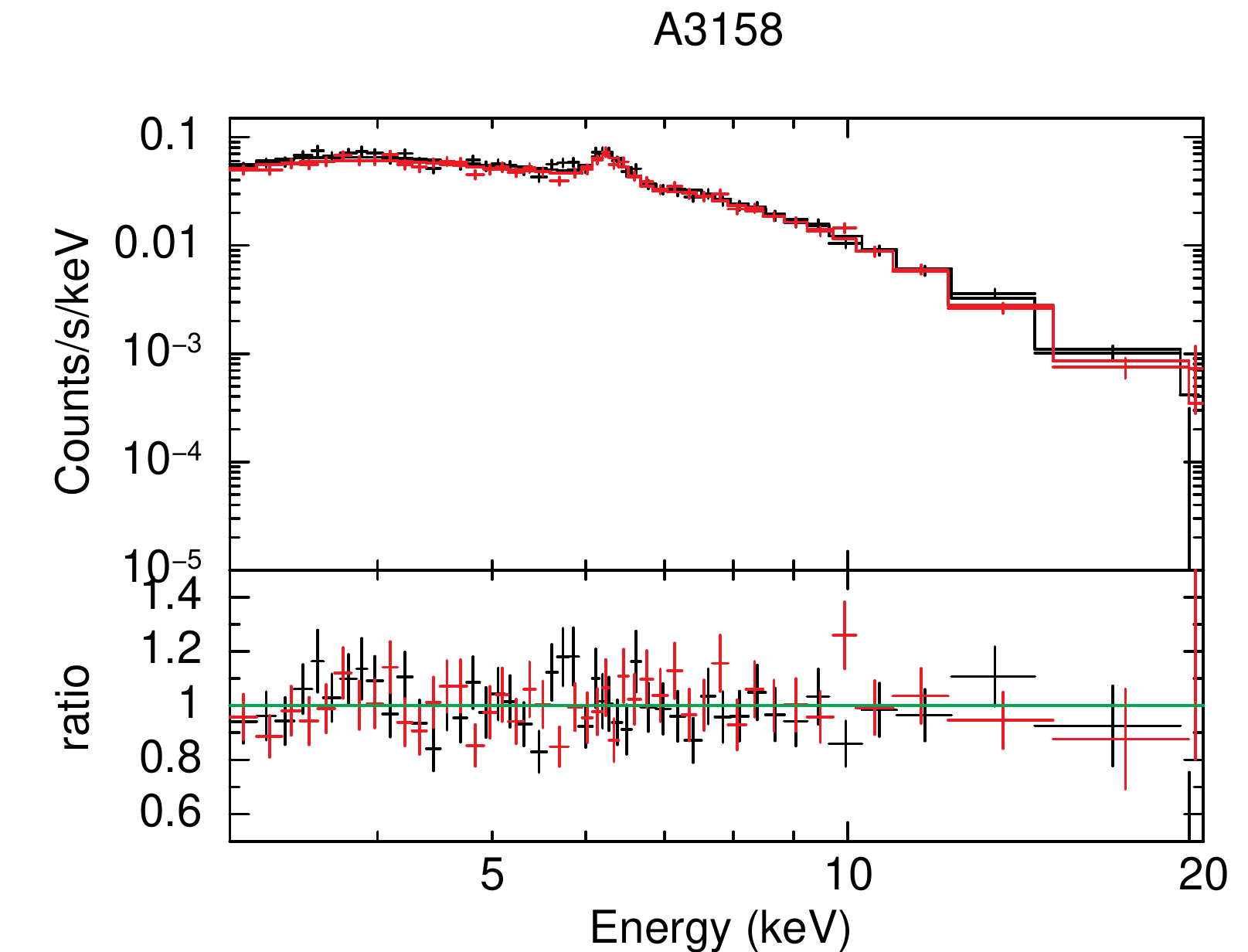} 
    \end{subfigure}
    \hfill
    \begin{subfigure}[t]{0.5\linewidth}
        \centering
        \includegraphics[width=0.72\linewidth, angle =270]{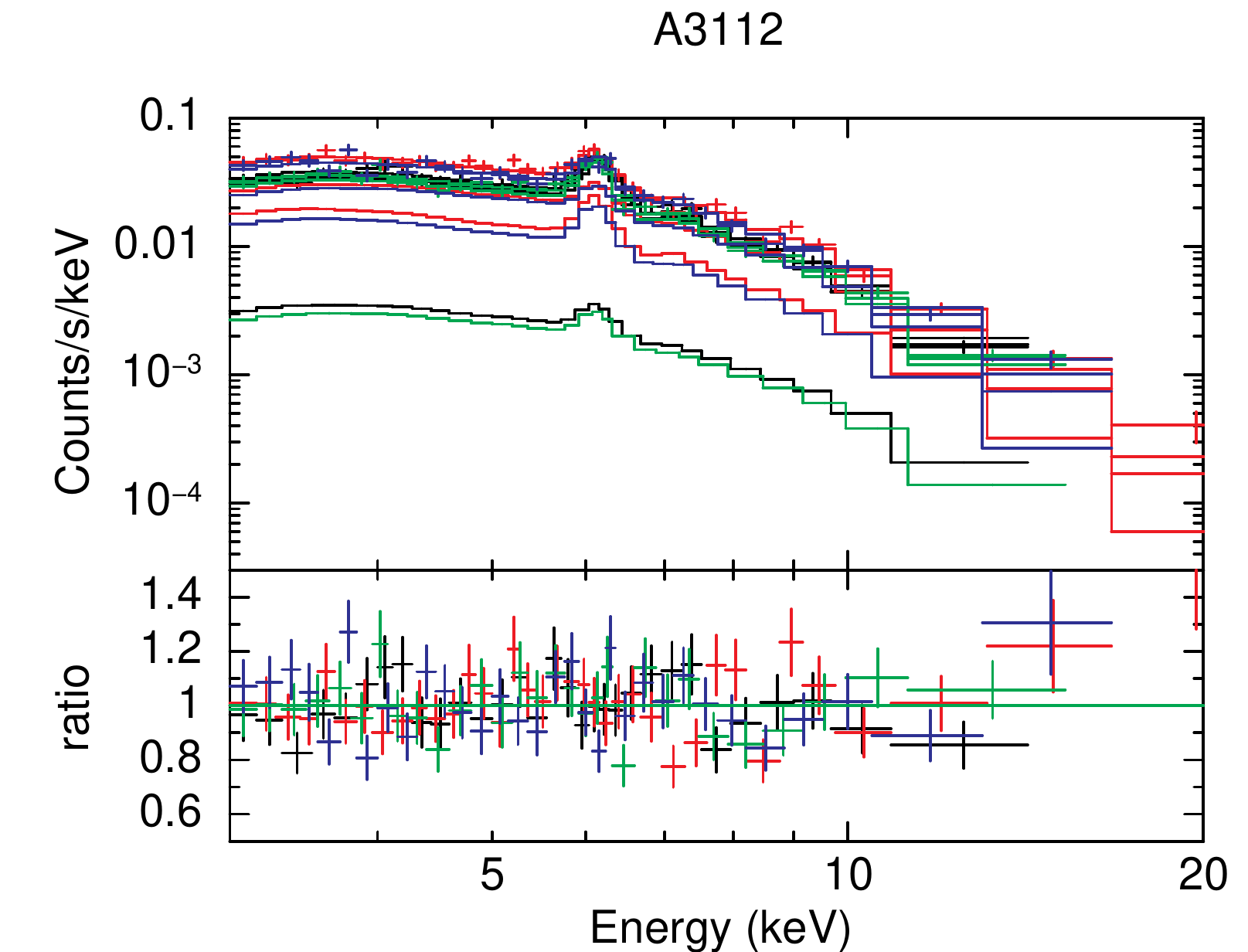} 
    \end{subfigure}
    \par\vspace{0.5cm}\par
    \begin{subfigure}[t]{0.5\linewidth}
        \centering
        \includegraphics[width=0.72\linewidth, angle =270]{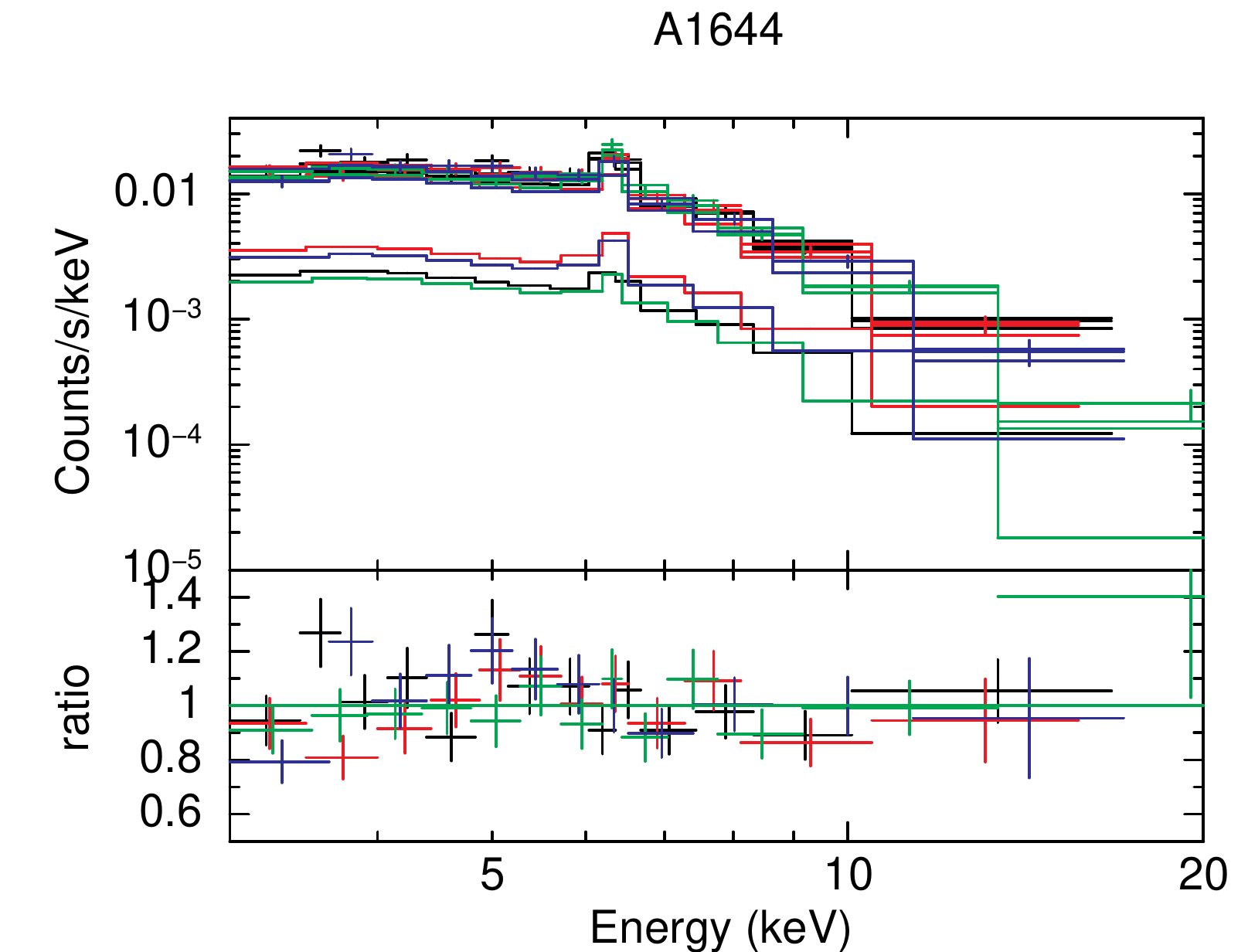} 
    \end{subfigure}
    \hfill
    \begin{subfigure}[t]{0.5\linewidth}
        \centering
        \includegraphics[width=0.72\linewidth, angle =270]{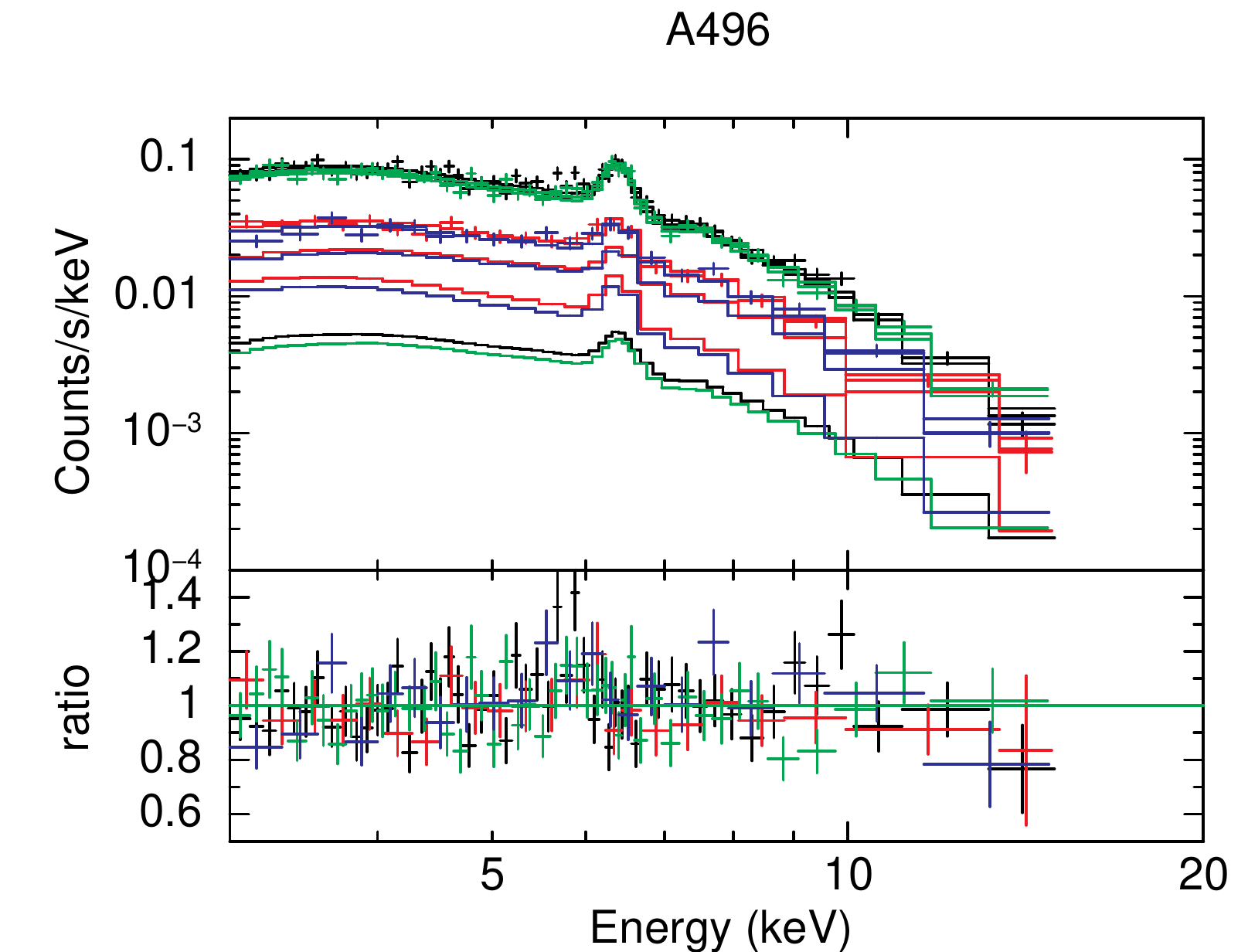} 
    \end{subfigure}
    \caption{Same as Figure~\ref{fig:spec1} but for clusters A3158, A3112, A1644, and A496.}
    
    \label{fig:spec2}
\end{figure*}

\subsection{NuSTAR Cross Talk}
\label{sec:proc:psfscat}

Galaxy clusters with cool cores have temperature profiles that decrease significantly towards the center of the cluster.
For cross-calibration studies, regions of near uniform temperature are needed to reduce different average temperature measurements due to bandpass effects \citep[e.g.,][]{Mazzotta}.
Our cluster sample includes six cool core clusters.
Due to the 58\arcsec\ half-power diameter of the \nustar\ point spread function (PSF), a non-trivial fraction of the emission originating in one part of the cluster will be recorded elsewhere on arcminute scales.
While we wish to ignore the 
bright central cool core emission, significant numbers of photons will be scattered into the annulus of interest, potentially biasing a temperature measurement there.
In order to account for the cross scattering or contamination in these clusters, 
we use knowledge of the energy-dependent PSF to create auxilliary response files (ARFs) that account for emission scattered from one region into another using the {\tt nucrossarf}\footnote{\url{https://github.com/danielrwik/nucrossarf}} code \citep[e.g.,][]{Potter, Tumer}.
For source emission contained in N extraction regions, {\tt nucrossarf} creates $N \times N$ ARFs, or cross-ARFs, to reflect the fraction of emission as a function of energy from each region that falls inside each of the $N$ regions.
The cross-ARFs are applied to $N$ sets of spectra extracted from a given region in {\tt Xspec}, with each cross-ARF for an origin region corresponding to a source number in {\tt Xspec}, and the model for that source is folded through the $N$ cross-ARFs and applied to all $N$ spectra.


An inner region was created around the core based on the cool core radius laid out by \cite{Schellenberger} as well as an outer annular region of radius 3.5\arcmin.
In principle, emission from outside the 3.5\arcmin\ outer radius is scattered in, but as the ICM surface brightness falls off quickly at these radii, we ignore this contribution 
as it typically contributes $<$1\% to the photons in the region; in any case, the temperature is generally similar and would not cause significant bias even if the emission were bright, which it is not. 
Inputs to {\tt nucrossarf} include the observation data paths, astrometric shifts to correct for offsets in both \nustar's absolute and relative (between FPMA and FPMB) astrometry, extraction regions, and a model of the source distribution inside the regions for which we use the \nustar\ background-subtracted and exposure-corrected image in the 3\textendash20~keV bandpass.
An identical analysis strategy was employed to analyze \nustar\ observations of the Perseus cluster, and as part of that work a study was done to evaluate {\tt nucrossarf}'s ability to model the outer PSF emission of point sources, finding accurate reconstructions on the order of a few percent \citep{Creech}.

For the six clusters with an inner exclusion region (like that shown in Fig.~\ref{fig:abell_0496}), from which spectra are extracted in both the inner and outer annular regions, the 3--20~keV spectral fits are shown in Figures~\ref{fig:spec1} \& \ref{fig:spec2}; these clusters have 4 spectra, presented in black, red, green, and blue.

For the 4 clusters with single regions (like that shown in Fig.~\ref{fig:abell3571}), the 3--20~keV spectral fits are shown in Figures~\ref{fig:spec1} and \ref{fig:spec2}; these clusters only have two spectra, presented in black and red.

\begin{deluxetable*}{lcccccc}
\scriptsize
\tablehead{ & Redshift & Gain Offset & $kT_c$ &  $kT_X$ & $kT_N$ & $kT_{Ni}$ \\ Cluster Name & z & keV & keV & keV & keV & keV}
\startdata
        RXC J1504 & 0.2172 & $-0.10 \pm 0.04$ & $9.81 ^{+0.80}_{-0.79}$ &  $6.40^{+0.20}_{-0.16}$ & $8.55 ^{+0.55}_{-0.48}$ & $5.93 ^{+0.09}_{-0.17}$  \\
        Abell 3571 & 0.0390 & $-0.09 \pm 0.02$ & $8.10 ^{+0.08}_{-0.08}$ & $6.36 ^{+0.06}_{-0.03}$ & $7.12 ^{+0.05}_{-0.10}$\\
        Abell 3558 &  0.0484 & $-0.05 \pm 0.03$ & $7.42 ^{+0.27}_{-0.28}$ & $5.51 ^{+0.08}_{-0.08}$ & $6.00 ^{+0.20}_{-0.20}$  & $6.23^{+0.15}_{-0.15}$ \\
        Abell 1651 & 0.0850 & $-0.09 \pm 0.04$ & $7.07 ^{+0.25}_{-0.25}$ & $6.09 ^{+0.12}_{-0.12}$ & $6.73 ^{+0.10}_{-0.07}$\\
        Abell 3391 & 0.0561 & $-0.10 \pm 0.07$ & $6.62 ^{+0.22}_{-0.22}$ & $5.54 ^{+0.13}_{-0.09}$ & $6.24 ^{+0.20}_{-0.20}$ \\
        Abell 1650 & 0.0838 & $-0.10 \pm 0.04$ & $6.43 ^{+0.10}_{-0.10}$ & $5.14 ^{+0.05}_{-0.05}$ & $6.55 ^{+0.10}_{-0.10}$ & $5.88 ^{+0.45}_{-0.15}$ \\
        Abell 3158 & 0.0592 & $-0.05 \pm 0.05$ & $6.01 ^{+0.10}_{-0.10}$ & $5.11 ^{+0.10}_{-0.08}$ & $5.79 ^{+0.11}_{-0.11}$\\
        Abell 3112 & 0.0753 & $-0.10 \pm 0.03$ & $5.45 ^{+0.12}_{-0.09}$ & $4.00 ^{+0.06}_{-0.04}$ & $5.57 ^{+0.20}_{-.020}$ & $4.59 ^{+0.10}_{-0.10}$ \\
        Abell 1644 & 0.0474 & $-0.06 \pm 0.05$ & $5.31 ^{+0.14}_{-0.13}$ & $4.61 ^{+0.19}_{-0.17}$ & $5.23 ^{+0.15}_{-0.15}$ & $5.24 ^{+0.10}_{-0.10}$ \\
        Abell 496 & 0.0331 & $-0.07 \pm 0.03$ & $5.18 ^{+0.07}_{-0.07}$ & $4.39 ^{+0.11}_{-0.08}$ & $5.40 ^{+0.15}_{-0.05}$ & $3.82 ^{+0.05}_{-0.02}$ \\
\enddata
\caption{\label{temps} The temperature measurements found for the cluster sample. $kT_C$ and $kT_X$ denote the \chandra\ and \xmm\ temperatures, respectively, measured in the outer region when fit in the 0.7\textendash7~keV band \citep{Schellenberger}.
$kT_{Ni}$ and $kT_N$ denote the \nustar\ temperatures measured in the inner and outer regions, respectively, when fit in the 3\textendash20 keV band with {\tt nucrossarf}. 
}
\label{tab:kts}
\end{deluxetable*}

The global temperatures as well as the inner region temperatures for the cool core clusters are presented in Table~\ref{temps}, along with the measurements from the same regions of these clusters with \chandra\ and \xmm\ from \citet{Schellenberger}. The \xmm\ temperatures are from a combined fit to all EPIC detectors, MOS1, MOS2, and PN.
\nustar\ temperatures reported for this sample are derived from joint spectral fits of data from both FPMA and FPMB.
For clusters without a cool core, a single temperature from one global region is found with the standard response files produced by {\tt nuproducts}.
For cool core clusters, fits were simultaneously determined in two regions, with the fit to the inner region simply providing a characterization of the emission scattered from there into the outer annulus of interest.

\section{Results} 
\label{sec:results}


\begin{figure*}
    \centering
    \begin{subfigure}[t]{0.48\textwidth}
        \centering
        \includegraphics[width=\linewidth]{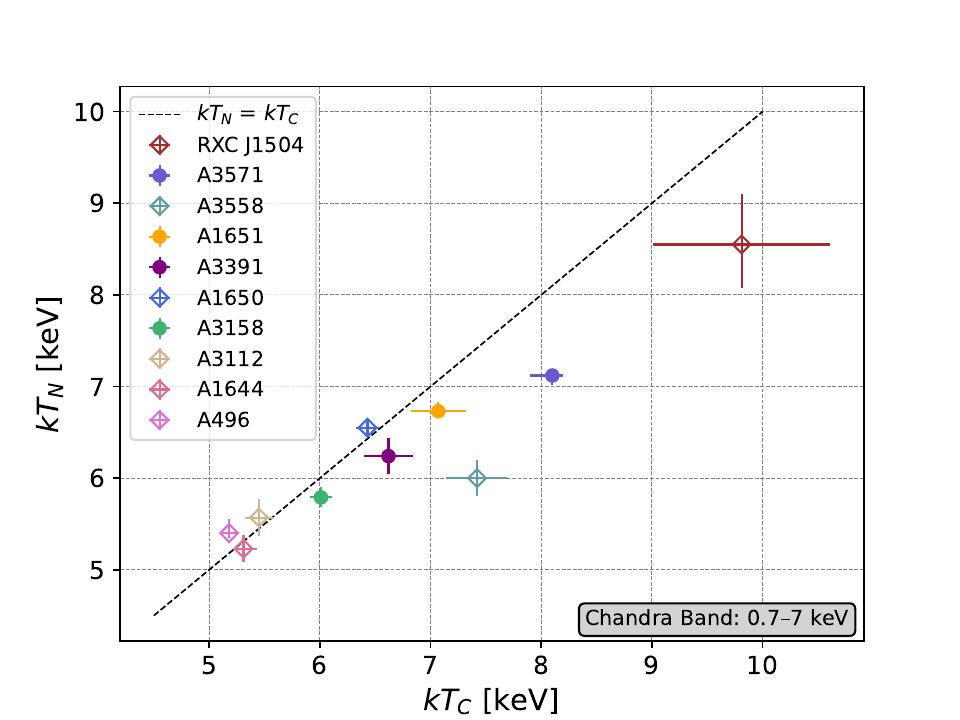}
    \end{subfigure}
    \begin{subfigure}[t]{0.48\textwidth}
        \centering
        \includegraphics[width=\linewidth]{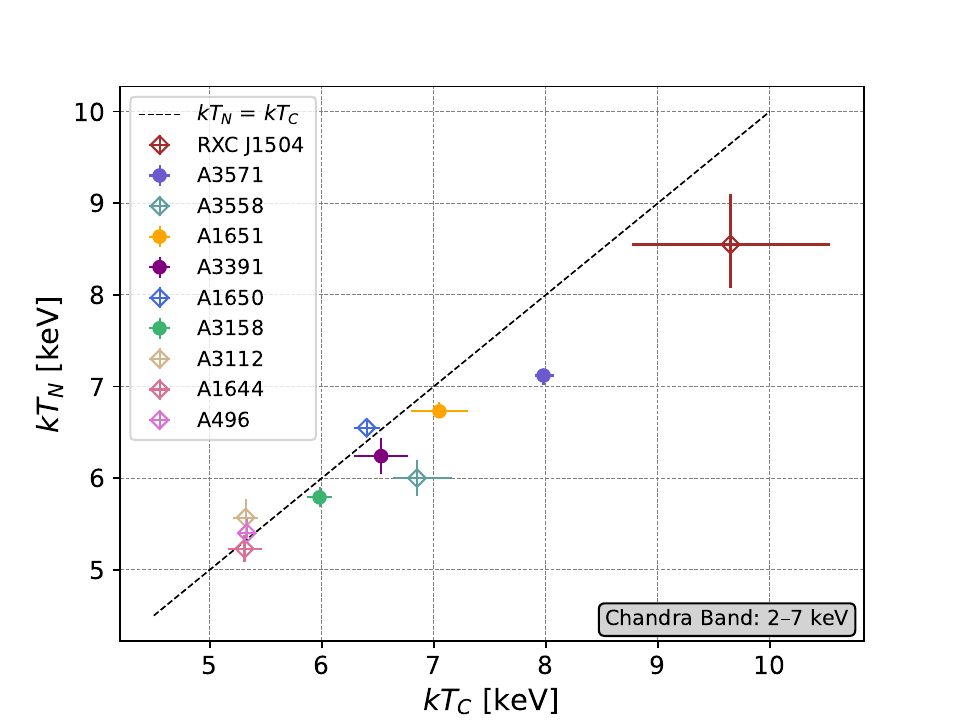}
    \end{subfigure}
    \caption {The \nustar\ temperatures $kT_N$ (y-axes) are compared to 
    those from \chandra\ \citep{Schellenberger}; the \chandra\ temperatures are measured in the 0.7--7~keV band  (left panel, values from Table~\ref{tab:kts}) and the 2--7~keV band (right panel, given in \citet{Schellenberger}).
    The clusters with the centers excluded  are indicated by an empty diamond symbol. \chandra\ temperatures are in general agreement regardless of the bandpass used to measure the temperature.}
    \label{fig:temp_plot}
\end{figure*}

Figure \ref{fig:temp_plot} presents comparisons between \nustar\ temperatures measured in the 3--20 keV band and \chandra\ temperatures measured in the 0.7--7 keV and 2--7~keV bands. 
These comparisons aim to reflect the differing results from each observatory for this relaxed cluster sample. 
Despite the differing bandpasses and relative sensitivity across those bandpasses the temperatures were derived from, good agreement is found for the lower temperature clusters in the sample ($5~{\rm keV}<kT<7~{\rm keV}$).
Only for hotter clusters does the agreement break down, with \chandra\ temperatures becoming systematically higher than the \nustar-measured ones. These results are unchanged if a different \chandra\ bandpass is considered, as the temperatures derived are consistent \citep{Schellenberger}.
This latter trend was previously seen for hotter, merging clusters \citep{Wallbank}, which are preferentially selected for by Guest Observers and thus make up the majority of galaxy cluster observations in the \nustar\ archive.
The uncertainties on the temperature measurements of RXC~J1504 are significantly larger due to the presence of a strong cool core.
In the case of \chandra, the hotter temperature and fainter emission (higher redshift) of the cluster results in less precision; for \nustar, the scattered cool core emission into the outer annulus creates a higher effective background, making a more precise measurement of the faint emission difficult.  RXC~J1504 is the highest redshift cluster in the sample, which makes the situation with scattered cool core emission worse for \nustar.

\begin{figure*}
    \centering
    \begin{subfigure}[t]{0.48\textwidth}
        \centering
        \includegraphics[width=\linewidth]{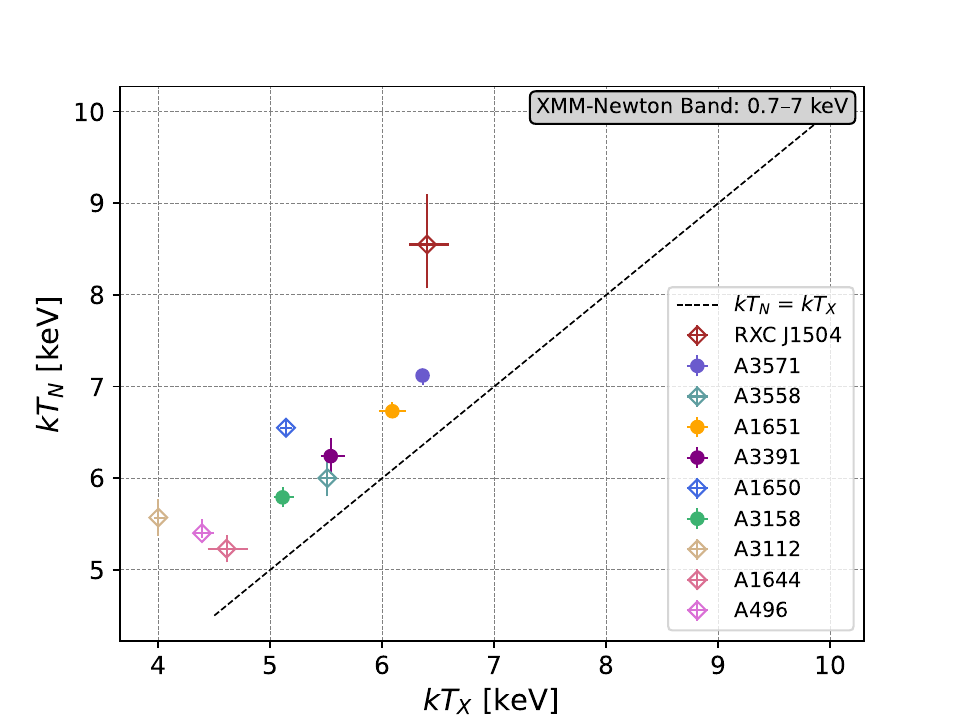}
    \end{subfigure}
    \begin{subfigure}[t]{0.48\textwidth}
        \centering
        \includegraphics[width=\linewidth]{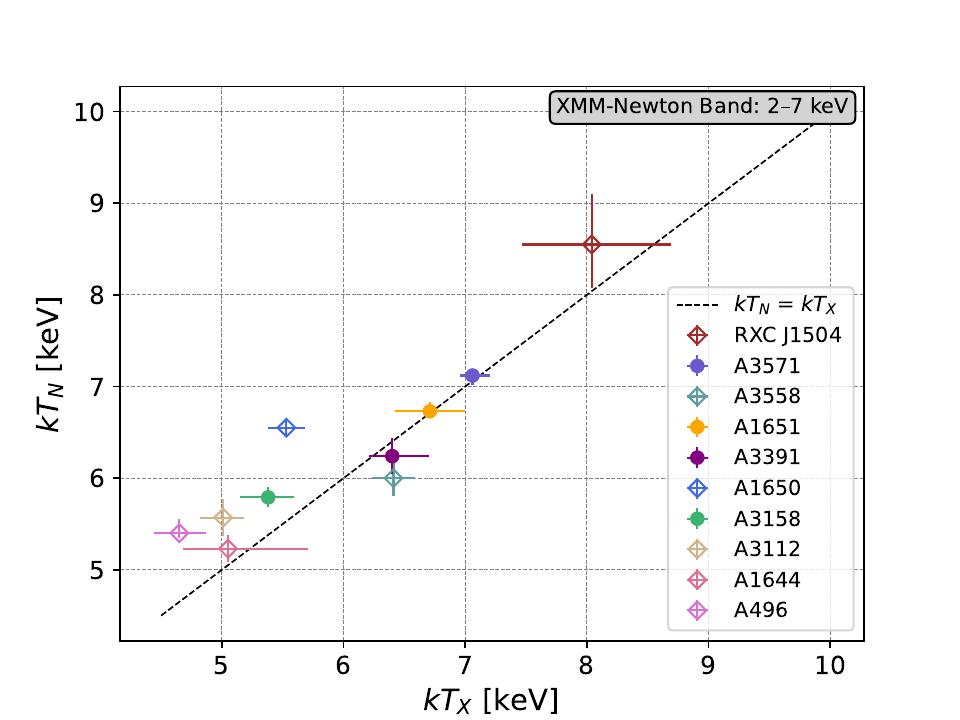}
    \end{subfigure}
    \caption {The \nustar-derived temperatures $kT_N$ compared to 
    those measured with \xmm\ \citep{Schellenberger}; the \xmm\ temperatures are measured in the 0.7--7~keV band  (left panel, values from Table~\ref{tab:kts}) and the 2--7~keV band (right panel, given in \citet{Schellenberger}).
    The clusters with the centers excluded  are indicated by an empty diamond symbol. \xmm\ temperatures change significantly depending on the bandpass used; while the 0.7--7~keV temperatures all fall below the \nustar\ estimates, this result is physically possible if there is significant cool gas spatially coincident present that \nustar\ spectra are not sensitive to.}
    \label{fig:temp_plot_xmm}
\end{figure*}

Figure \ref{fig:temp_plot_xmm} presents temperature comparisons between \nustar\ and \xmm, also derived in broad (0.7--7~keV) and hard (2--7~keV) energy bands.
\nustar\ temperature measurements exceed \xmm\ broadband temperatures on the order of 15\% for most systems, although in a few cases the discrepancy is larger, namely RXC~J1504, A1650, and A3112. 
Compared to Figure~\ref{fig:temp_plot}, the \nustar\ temperatures are in worse overall agreement compared to those from \xmm, although the direction of the difference is more in line with physical expectations; \nustar's harder bandpass and sensitivity should weight hotter components more when multi-temperature distributions are present, leading to a higher average (measured) temperature estimate.
However, given the lack of evidence for multi-temperature structure in the \nustar\ spectra (see Section~\ref{sec:results:bandpass} below), the difference would need to be due to soft components with nearly negligible emission in \nustar's bandpass.

Also, the outliers do not show a systematic trend here as they do in Figure~\ref{fig:temp_plot}; the three most discrepant clusters span the entire range of cluster temperature sampled.
This result is consistent with recent results on the internal cross calibration of instruments onboard \xmm, which finds a cluster-to-cluster scatter in the determination of a scaling factor meant to bring instrument measurements into agreement \citep{NevMol23}.
In other words, the source-to-source variation seen in Figure~\ref{fig:temp_plot_xmm} appears to be related to calibration for \xmm. 
Ignoring the scatter, a simple bias appears able to reconcile the measurements, although the cause of such a bias is unclear.

A systematic trend is more apparent when hard band \xmm\ temperatures are compared to \nustar's (Fig.~\ref{fig:temp_plot_xmm}, right panel), with the temperatures showing better agreement as the temperature of the system increases.
However, this consistency directly contradicts non-cluster studies, where simultaneous \xmm\ and \nustar\ observations show strong and systematic disagreement when fit with the same spectral model at hard energies (see Section~\ref{sec:disc:hardband} for a discussion).
Applying a correction to the \xmm\ EPIC hard band effective areas, as is now the default in SAS (versions 22 and later), would drive the measured hard band temperatures lower such that the trend would likely be similar to that seen using the broadband.

Temperature measurements could potentially be biased by details of the modeling, such as by an incorrect redshift or gain, abundance, or more complex thermal structure (either internal to the spectral extraction region or along the line-of-sight) 
A small gain shift on the order of what was measured in Table~\ref{tab:kts} ($\sim$100~eV) has previously been seen in other cluster observations as well as some AGN, and it is could be a result of a minor issue with the low energy gain calibration; instrumental lines and supernova remnant observations set the gain scale at higher and lower energies, respectively. 
However, apparent gain shifts could also be induced by the application of an incorrect model.
The presence of significant cooler, higher metallicity gas would more emphasize the 6.7~keV line relative to the 6.9~keV line, pushing the Fe complex line centroid to lower energies.
In any case, gain shifts of this order have negligible impact on continuum-based estimates of temperature for low redshift ($z<0.1$) clusters when the redshift is left as a free parameter.
However, at higher redshift, the multiplicative effect of a redshift change versus an additive gain offset can affect the measured temperature; there is only one cluster (RXC~J1504) where this effect could be important, but the measured temperature only differs by $\Delta kT \sim 0.1$~keV, much less than the statistical uncertainty in this case and on the order of statistical uncertainties for the other clusters in the sample.
Thus, the true presence of a gain shift will not affect our results.
Similarly, at \nustar\ energies, the impact of emission lines on the continuum shape---potentially biasing the continuum-based temperature measurement---is isolated to the $\sim$6.4--7~keV Fe complex.
Because these lines are not resolved, a simple rescaling of the overall abundance, assuming solar abundance ratios, is sufficient to account for them and unlikely to lead to mismodeling in a way that could bias the temperature.
Multi-temperature structure within the spectral extraction region, on the other hand, could play a larger role in biasing measurements; this possibility is investigated in Section~\ref{sec:results:bandpass}.

%
%
%

\begin{deluxetable}{lcccccc}
\tabletypesize{\scriptsize}
\tablewidth{0pt}
\tablehead{Cluster Name & $kT_{N,(3\textendash8)}$ & $kT_{N,(7\textendash11)}$ & $kT_{N,(11\textendash20)}$ & $kT_{N}$ \\ & keV & keV & keV & keV}
\startdata
        RXC J1504 & $7.2 \pm 2.0$ &  $>8.8 $ & $12.2 \pm 25.6$ & $8.6 \pm 0.6$  \\
        Abell 3571 & $7.7 \pm 0.4$ & $6.6 \pm 0.5$ & $7.1 \pm 0.8$ & $7.1 \pm 0.1$   \\
        Abell 3558 & $5.6 \pm 0.8$ & $7.7 \pm 1.8$ & $4.7 \pm 1.9$ & $6.0 \pm 0.1$ \\
        Abell 3391 & $6.2 \pm 0.8$ & $6.3 \pm 1.4$ & $8.1 \pm 4.1$ & $6.2 \pm 0.2$ \\
        Abell 1650 & $6.9 \pm 0.9$ & $6.7 \pm 1.4$ & $6.5 \pm 2.2$ & $6.6 \pm 0.1$ \\
        Abell 3158 & $5.9 \pm 0.5$ & $6.6 \pm 0.7$ & $5.2 \pm 1.2$ & $5.8 \pm 0.1$\\
        Abell 3112 & $5.2 \pm 0.7$ & $5.6 \pm 1.9$ & $6.8 \pm 4.9$ & $5.6 \pm 0.2$ \\
        Abell 1644 & $5.9 \pm 1.3$ & $4.1 \pm 1.0$ & $>3.9 $& $5.2 \pm 0.3$ \\
        Abell 1651 & $7.1 \pm 0.6$ & $6.2 \pm 0.7$ & $7.1 \pm 3.8$ & $6.7 \pm 0.2$\\
        Abell 496 & $6.7 \pm 1.4$ & $4.9 \pm 1.4$ & $8.6 \pm 17.0$ & $5.4 \pm 0.1$\\
\enddata
\caption{The temperature measurements found for the cluster sample. $kT_{N,({\rm range})}$ denotes the \nustar\ temperatures measured in the bandpass range, $kT_N$ denotes the \nustar\ temperatures measured in the 3\textendash \textbf{11} keV band.
}
\label{tab:bandpass}
\end{deluxetable}

\begin{figure*}
    \centering
    \includegraphics[width=\linewidth]{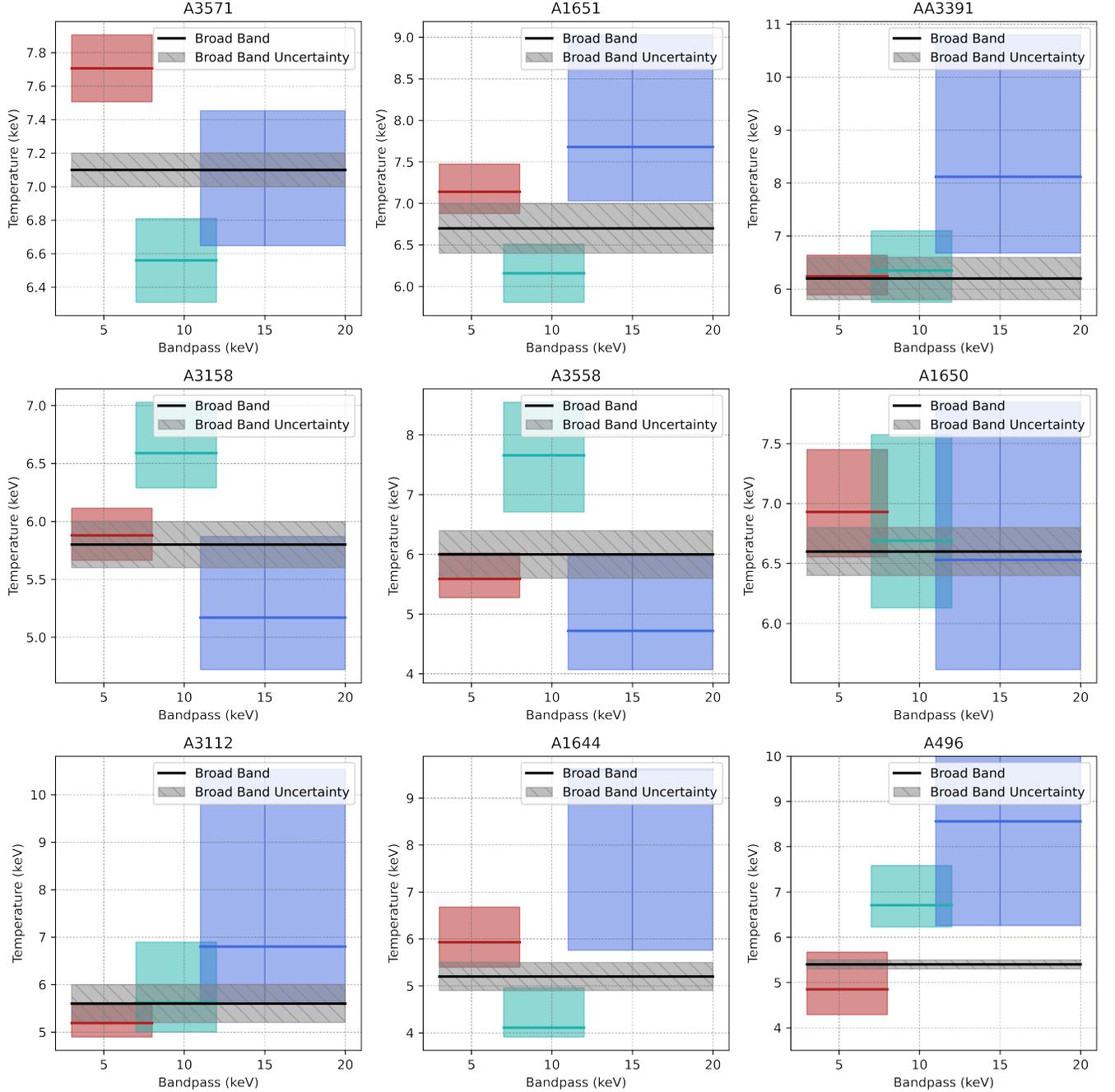}
    \caption {Observed temperature measurements in relation to bandpass range for each cluster in the relaxed cluster sample. The bandpass ranges include 3\textendash8 keV (red), 7\textendash12 keV (green), and 11\textendash20 keV (blue). The bandpass ranges are compared to the overall broad band temperature found for each cluster.}
    \label{fig:bandpass_comp}
\end{figure*}

\subsection{Changing the Bandpass}
\label{sec:results:bandpass}

In order to identify the effect of multiphase gas within the measurement region, each cluster was fit in multiple overlapping spectral bands to understand how changing the bandpass affects the measured temperature. 
Due to worsening statistics, the abundance and gain offset were frozen and set to the best fit parameters from the full band spectral fitting. 
The abundance was left as a free parameter for the bandpasses where the Fe line is present and could constrain it, otherwise it was fixed to the value found in the full band fit.
The bandpass was fit in two broader ranges first, namely from 3--11 keV and 11--20 keV energies.
Given the large uncertainty in the hard bandpass, the data were further separated at soft energies into two narrower bandpasses: 3--8 keV and 7--12 keV. 

Temperatures in each band, including from the full 3--20~keV band, are reported in Table~\ref{tab:bandpass} and compared for each cluster in Figure~\ref{fig:bandpass_comp}.
Although the uncertainties are larger in these narrow bands, a low significance systematic trend could still reveal itself as an increase in the measured temperature as the bandpass energy increases, which is what is seen in A496 and to a lesser extent, in A3112.
Other systems show more random variations or the opposite trend, which would not be expected from a physical distribution of temperatures.
A3112 is known to have an apparent hard (and also soft) excess, which has been interpreted in the past as being due to inverse Compton emission \citep{Bonamente,Abell3112}; these results do not support that interpretation, but they are not inconsistent with the kind of multi-temperature structure that can masquerade as non-thermal emission \citep[e.g.,][]{Valinia_2000,Tanaka}.
However, given the size of the uncertainties, 
the null hypothesis that the fluctuations are simply randomly observed deviations from a single, non-bandpass-dependent temperature cannot be rejected.

\begin{figure}
    \centering
    \includegraphics[width=\linewidth]{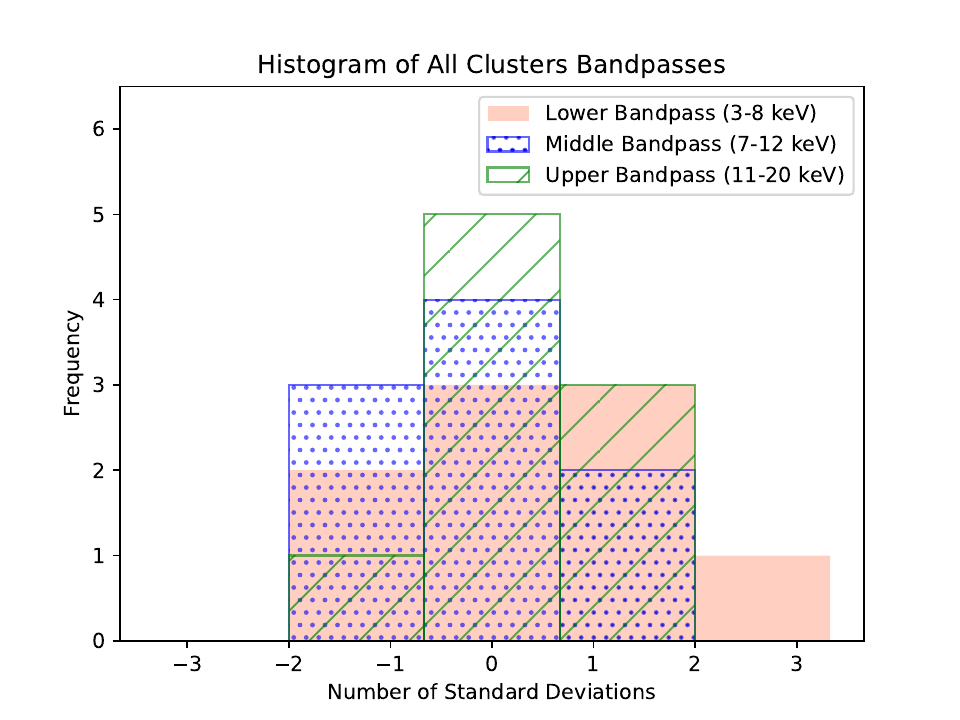}
    \caption {Number of standard deviations of the bandpass range (lower, middle, upper) in relation to the broad band pass in relation to amount of temperature values. The amount of standard deviations is found using (\ref{gaussian}).}
    \label{fig:histogram}
\end{figure}

To better understand if the differences between the broadband and narrower bandpass-determined temperatures are statistically significant, the scatter in the measurements were compared to the expected scatter assuming Gaussian fluctuations.
We recenter the measured temperature $T_i$ by the full bandpass-determined temperature $T_{\rm broad}$ and scale by the narrow bandpass temperature uncertainty $\sigma_i$:
\begin{equation}
    \Delta \chi_T = \frac{T_i-T_{\rm broad}}{\sigma_i}\, .
    \label{gaussian}
\end{equation}
If deviations from the broad bandpass temperature are due entirely to statistical fluctuations, we would expect the distributions of $\Delta \chi_T$ to be centered on zero with a standard deviation of unity, which is indeed what is seen (Figure~\ref{fig:histogram}). The mean ($\langle \Delta \chi_T \rangle$) and standard deviation ($\sigma$) were found for each bandpass; 3--8 keV ($\langle \Delta \chi_T \rangle$ = 0.58, $\sigma$ = 1.06 keV), 7--12 keV ($\langle \Delta \chi_T \rangle$ = -0.21, $\sigma$ = 1.20 keV); 11--20 keV ($\langle \Delta \chi_T \rangle$ = 0.27, $\sigma$ = 0.59 keV).

Given the lack of evidence that would point toward the existence of multi-temperature structure (i.e., systematically increasing temperatures as the energy range of the bandpass increases), we do not rigorously pursue multi-temperature fits to the modest signal-to-noise spectral in this sample.
The application of simple two temperature models ({\tt APEC}+{\tt APEC}) when fitting the spectra result in one of the components typically having an extremely high or low temperature with low fractional flux---in essence, the second component attempts to fit some low significance feature at either end of the bandpass instead of adjusting the overall continuum shape to better fit the spectrum globally \citep[e.g.,][]{NuComa}. 

The unphysical nature of this component demotivates its addition to the model, even if the fit statistic in some cases is technically improved.
Therefore, we do not consider the potential impact of multi-temperature structures, as such an impact should be minimal if not negligible. 


\subsection{Considering a Temperature Distribution}
\label{sec:results:tdist}

Despite the lack of evidence for a strong influence of multi-temperature gas inside our regions from Section~\ref{sec:results:bandpass},
the gas is not expected to be truly isothermal.
Theoretical \citep[e.g.,][]{Rasia2014} and observational \citep[e.g.,][]{Lovisari2024chexmate} work shows that within $R_{500}$, the temperature can vary from one region to another, at the same projected radius, by 10--30\%.
Instruments with different bandpass sensitivities will weight this distribution differently, with \chandra\ and \xmm\ more sensitive to the lower temperature part of the distribution, while \nustar\ will more heavily weight the higher temperature part.
When a temperature distribution is fit with a single temperature model, the measured temperature will reflect this weighting.

For our simple \nustar\ spectra---the four clusters without excluded cores---we also fit them with the {\tt gadem} model in {\tt XSpec}, which produces a spectral model with a Gaussian distribution of temperatures.
The ATOMDB database was used to interpolate the model ({\tt switch = 2}), and the $n_H$ parameter was fixed to a low value since its effect is negligible.
We first explicitly fit for the distribution width, $\sigma_T$, to see if the spectra themselves were better fit by such a model.
In these four cases, the best fit value of $\sigma_T$ was consistent with zero at the 90\% level, with upper limits of $\sigma_T \lesssim 0.5$~keV.
The mean temperatures in all cases remain consistent with the original single temperature values found.
This result is not surprising, as the {\tt gadem} model, even for large fractional $\sigma_T/T$, deviates only slightly from a single {\tt APEC} model in terms of continuum shape.
However, this result still suggests that multi-temperature gas is not significantly biasing our temperature measurements.

We also fixed $\sigma_T$ to 30\% of the single {\tt APEC} best-fit temperature and fit for the mean temperature, to see how the temperature would typically be biased assuming the underlying gas follows such a distribution.
The temperatures using the {\tt gadem} model, with fixed nonzero $\sigma_T$, shift only very slightly lower than our single {\tt APEC} measured temperatures from Table~\ref{tab:kts}, all still well within the 90\% uncertainty envelope.
This negligible change is consistent with the difference in temperatures found from simulations of \nustar\ spectra, using the {\tt gadem} model but fit with a single temperature {\tt APEC} model; a $\sigma_T$ of 50\% increases the measured temperature by only 3.5\% compared to the mean.

\section{Discussion} 
\label{sec:disc}

\subsection{Comparison of the Measurements from the Three Observatories}
\label{sec:disc:basic}

Our main results are shown in Figures~\ref{fig:temp_plot} and \ref{fig:temp_plot_xmm}, which compare our \nustar\ temperature measurements with those for \chandra\ and \xmm\ from the same regions reported in \citet{Schellenberger}.
In both comparisons, we find systematic disagreements between the measured temperatures.
While the \chandra\ and \nustar\ temperatures agree for the lower temperature systems, in higher temperature ones the \chandra\ temperature skews higher.
In contrast, the \nustar\ temperatures are found to be systematically higher than those from \xmm, with no clear trend with temperature.

As discussed in Section~\ref{sec:results:tdist} and in the following sections, the presence of multiple temperature components, when fit to a single temperature model, will skew the average temperature depending on the instrument's bandpass and the relative energy sensitivity across it.
\nustar's harder bandpass and peak sensitivity around 10~keV should therefore produce higher average temperature measurements than either \xmm\ or \chandra\ in the presence of multi-temperature gas.
This result is exactly what is seen when \nustar\ and \xmm\ are compared (Figure ~\ref{fig:temp_plot_xmm}).
While contradictory calibrations could be the culprit for this disagreement, a physical cause cannot be ruled out.
In particular, cool 1--2~keV gas would produce very little relative emission in the \nustar\ bandpass, while appearing at the peak of \xmm's sensitivity.
The greater flux at soft energies would create a steeper spectrum in the \xmm\ bandpass, leading to a lower single temperature measurement.

The opposite situation is seen in the \nustar\\chandra\ comparison (Figure ~\ref{fig:temp_plot}), and the existence of hotter \chandra\ temperatures cannot be explained with a physical cause such as multi-temperature gas.
The only viable explanation is a calibration discrepancy between the two observatories.
For that reason, we focus on this comparison, and its possible explanation in the remainder of the discussion.

\begin{figure}
    \centering
    \includegraphics[width=1.1\linewidth]{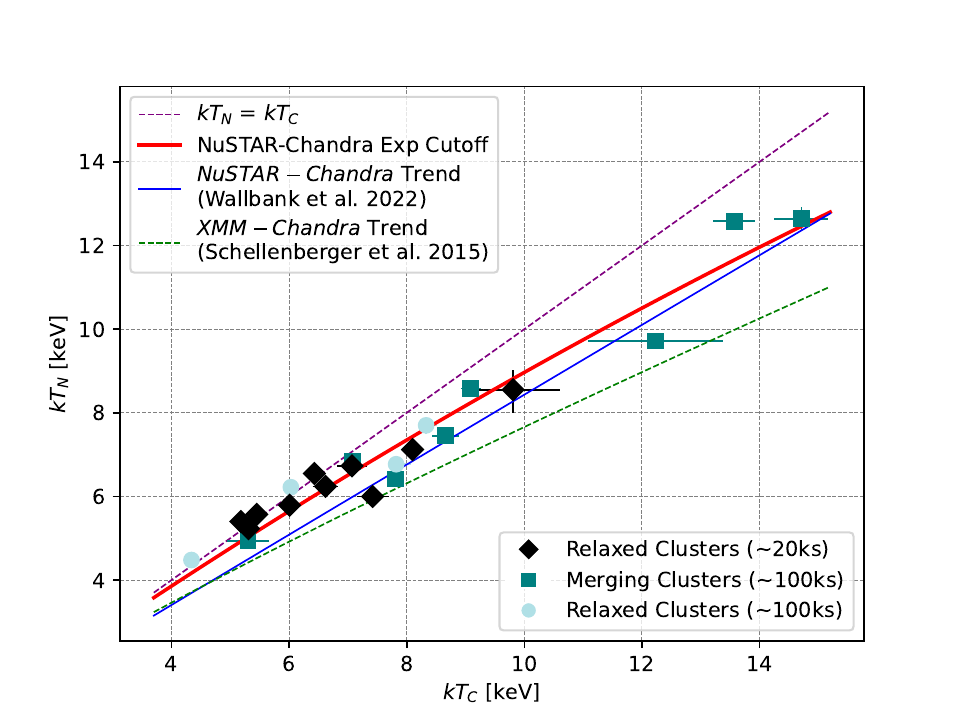}
    \caption {Temperature comparisons between \chandra\ and \nustar\ for 3 different samples. The relaxed cluster sample temperatures (this work, black diamonds) are shown alongside measurements of 4 deeper observations of relaxed clusters (light green circles) and a merging cluster sample \citep[dark green squares][]{Wallbank}. The best fit line to the latter data points from that work is drawn in blue, while the relation between \chandra\ and \xmm\ is given by a green dashed line \citep{Schellenberger} (the ``ACIS-Combined XMM Full" model is shown, measured in the 0.7--7 keV energy band, with \xmm\ temperatures in place of \nustar\ temperatures on the vertical axis).
    In red, we show the best-fit relation to all data for a model that assumes systematic uncertainties of the \chandra\ effective area above 2--3~keV (see Section~\ref{sec:disc:hardband} for details).
    }
    \label{fig:crosscalib}
\end{figure}

\subsection{\chandra/\nustar\ Cross Calibration}
\label{sec:disc:cc}

To understand the relationships between \nustar\ and \chandra\ temperatures, we include our results with other temperature measurements of galaxy clusters (Figure \ref{fig:crosscalib}). 
The temperatures presented here, based on shallow observations of 10 relaxed clusters are plotted with black diamonds. 
Using dark green squares, we show the temperature measurements from \cite{Wallbank}, which consist of merging systems due to their over-representation in \nustar's archive. We also include temperature measurements for four relaxed clusters with deeper \nustar\ observations \citep[A2199: $kT=4.48$ keV, A1795: $kT=6.22$ keV, A478: $kT=6.77$ keV, \& A2029: $kT=7.70$ keV;][2025 private communication]{Potter} and \chandra\ observations \citep[A2199: $kT=4.34$ keV, A1795: $kT=6.03$ keV, A478: $kT=7.82$ keV, \& A2029: $kT=8.33$ keV;][2025 private communication]{Potter} from regions extending out to 2.53\arcmin--3.80\arcmin.
For reference, we provide relations expected for equality (dashed purple line), the empirical \chandra--\xmm\ relation \citep[dashed green line;][]{Schellenberger}, and the best linear fit to the merging cluster sample \citep[solid blue line;][]{Wallbank}.
While the latter relation holds up well for clusters with $kT \gtrsim 7$--8~keV, it systematically underpredicts the \nustar\ temperature of cooler clusters (regardless of sample).
The change in behavior on either side of the $kT \sim 7$~keV divide hints at the nature of the discrepancy.
Despite temperature inconsistencies between \chandra\ and \xmm, \cite{Schellenberger} found that both observatories reported temperatures in better agreement with each other when only the hard band ($E>2$~keV) data were used, suggesting the calibration discrepancy originates at lower energies, where quantitative differences were also found.
The \nustar--\chandra\ discrepancy in hot clusters, however, challenges that hypothesis, since the temperature will be driven by measurements at hard energies (coincidentally in the overlapping bandpass $\sim$3--10~keV).
For cooler clusters, the \chandra\ measurement will be driven by statistics at softer energies, since the hard flux is rapidly declining and the $E>2$~keV effective area is relatively small.
The near 1:1 agreement at these lower temperatures suggests a more consistent cross-calibration between \nustar\ and \chandra's soft band.
In Section~\ref{sec:disc:hardband}, we investigate the impact small modifications to \chandra's hard band effective area would have on measured temperatures.

\subsection{Impact of Multi-temperature Components in Spectral Extraction Regions}
\label{sec:disc:multit}

Given the large regions used, a single temperature model is really an average temperature of the photons detected by the instrument: an average weighted by the telescope's sensitivity (bandpass and effective area).
The presence of higher temperature gas in a region will produce more hard than soft photons, due to bremsstrahlung's flatter low energy slope and higher energy exponential cutoff, so instruments with more sensitivity at higher energies should generally measure higher temperatures.
While \chandra\ and \xmm's bandpasses are comparable, their relative effective areas differ, particularly at higher energies, such that one would expect a higher temperature from \xmm.
The opposite situation is in fact the case, and even within \xmm\ the higher throughput instrument at hard energies---EPIC-pn---returns lower temperatures than the MOS instruments.
The much harder bandpass of \nustar\ should yield even hotter temperatures in the presence of multi-temperature gas, yet they also fall below those measured with \chandra.

If the emission is highly non-isothermal, the different bandpass sensitivities would potentially return different temperatures for the gas in a single temperature model fit.
Simulations using the {\tt gadem} model in {\tt XSpec}, which creates a Gaussian distribution of temperature components with a mean value and width $\sigma_T$, confirm this expectation.
As $\sigma_T$ increases, \chandra\ and \xmm\ temperatures (fit to a single $T$ APEC model) drop, while \nustar\ temperatures rise; when $\sigma_T$ is 50\% of the mean temperature in the simulations, the best-fit single temperatures differ from the mean by modest amounts (a drop of $\sim$10\% for both \chandra\ and \xmm\ and an increase of $\sim$3.5\% for \nustar).
This trend is true when the full energy range of each instrument is used.
We also performed the simulations in \nustar's soft (3--9~keV) and hard (9--25~keV) bands and find that the soft band temperature is very slightly lower than the simulated mean value while the hard band temperature is almost 10\% higher.
For the \nustar\ spectra we consider here, these systematic biases are less than the statistical uncertainty and suggest that \nustar\ measurements should provide accurate temperature estimates even in the presence of multi-temperature components.

\subsection{Impact of Hard Band Effective Area on Cross-calibration Trend}
\label{sec:disc:hardband}

The better agreement between \nustar\ and \chandra\ in clusters with temperatures below $\sim$7~keV suggests that \chandra's soft ($E\lesssim 3$~keV) calibration is more consistent with that of \nustar's, given that most photons detected by \chandra\ will fall in this energy range.
At higher temperatures, a higher fraction of emission ends up at $E>3$~keV energies; if there is a calibration offset
of \chandra's effective area at high energies, that could explain the divergence in temperature measurements for hotter gas.
A similar argument for \nustar\ is less easily made, as the emission is concentrated in the 3--10~keV band regardless of temperature.

A gradual, smooth increase in \chandra's effective area with energy above 2--3~keV is one possible modification, although there is no a priori form it must take.
However, a modification of this form has been proposed for \xmm's effective area---and is now applied by default in the latest versions of \xmm\ SAS (version 22.0 and later)---based on achieving a consistent cross-calibration between \xmm\ EPIC-pn and \nustar\ spectra of bright point sources, which are often simultaneously observed by both observatories and jointly fit\footnote{\url{https://xmmweb.esac.esa.int/docs/documents/CAL-TN-0230-1-3.pdf}}.
Given the better agreement of \xmm\ and \chandra\ temperatures when only the $E>2$~keV data are fit \citep{Schellenberger}, it is not unreasonable to expect a similar effective area modification to lead to better agreement between \chandra\ and \nustar\ temperature measurements as well.

We adopt the form of this modification (shown in Figure~7 of the \xmm\ Calibration Technical Note) and apply it to \chandra's effective area.
To assess how temperature measurements would be affected by this change, we simulate \chandra\ spectra (with {\tt XSpec}'s {\tt fakeit} routine) using a ``correct" (modified) effective area that are then fit with the unmodified, nominal effective area.
Absorbed single temperature {\tt APEC} models, spanning the range $4~{\rm keV} < kT < 14$~keV, are used to generate the simulated spectra; we use appropriate secondary parameters for this sample: $n_H = 10^{21}$~cm$^{-2}$, an abundance relative to solar of 0.3, and $z=0.05$.
The normalization is also set to produce spectra with roughly comparable signal-to-noise for our clusters.
Of course, the detailed properties from one cluster to another vary, but our goal is only to provide a toy model.

Because there is no reason to expect the \chandra\ effective area needs to be modified in exactly the same way as the \xmm\ effective area, we use this correction---obtained from the \xmm\ CALDB---as a starting point.
In our simulations, we vary both the overall size of the correction---multiplying the entire correction curve by a factor between 0 and 2---and by considering corrections that begin at lower energies.
The \xmm\ correction curve begins to modify the effective area at energies only above $\sim$3~keV, so to investigate the effect of corrections that start to appear at lower energies, we also perform simulations with the correction curve shifted to lower energies by either 0, 1~keV, or 2~keV.
Thus, we explore a family of effective area correction curves for which spectra with temperatures ranging between 4~keV and 14~keV are generated for a given scale factor and energy shift.

Regardless of how much we scale the correction by, or whether we shift the correction curve to lower energies, 
the trend between the input and best-fit temperatures can be fit by a consistent function.
We found that the simulations were well described by a generic exponential cutoff of the form
\begin{equation}
    kT_{\rm true} = kT_{\rm obs} e^{-kT_{\rm obs}/(kT_{\rm cut})}\, ,
\end{equation}
where $kT_{\rm obs}$ is the best-fit, or measured, temperature using a mismatched ARF, $kT_{\rm true}$ is the temperature input to the simulation, and $kT_{\rm cut}$ is a cutoff temperature and free parameter chosen to match the simulation results.
The unorthodox definition of the equation with respect to the observed and true temperatures allows the measured \chandra\ temperature ($kT_{\rm obs}$) to naturally appear on the x-axis, for easy comparison with previous work \citep[e.g.,][]{Schellenberger,Wallbank}.
Although we do not fundamentally know whether \nustar\ temperatures are more accurate,
we assume they are for this exercise.

To account for an additional systematic offset, due to bandpass weighting of projected emission or some other cross-calibration discrepancy, we also include a scaling factor $B$ when fitting to the measurements:
\begin{equation}
    kT_{Nu} = B \cdot kT_{Ch} \exp(-kT_{Ch}/kT_{\rm cut})\, .
\end{equation}
We find $kT_{\rm cut}=91\pm41$~keV and $B=1.00\pm0.05$.
The exponential form is in good agreement with the data, as is shown by the red line in Figure~\ref{fig:crosscalib}. 
There is little random scattering from \nustar\ and \chandra\ temperatures about the relation.
Compared to our spectral simulations, this value of $kT_{\rm cut}$ requires a greater than 2$\times$ 
increase to the \xmm\ effective area correction curve, which was at the top of our simulated range, when no energy shift is applied to the curve.
When the correction curve is shifted to lower energies by 1 or 2~keV, this cutoff temperature corresponds an overall scaling of the correction curve by a factor of 1.8$\times$ or 1.0$\times$, respectively.
In each case, this scaling corresponds to an increase in the \chandra\ effective area of 2\%/1\% at 3~keV, of 5\%/6\% at 5~keV, and 8\%/12\% at 7~keV.
Since the energy dependence of the correction is arbitrary---based on an empirical correction to \xmm's effective area---the exact values of these increases are not of interest and are beyond the ability of our data to constrain.
It is worth noting, however, that a $\sim$5\% increase in \chandra's hard band effective area around 5~keV is adequate to explain the observed discrepancy.
A more detailed investigation of potential modifications to the hard band effective area of \chandra\ is beyond the scope of this work.

\section{Conclusions} 
We used \nustar\ observations of 10 relaxed galaxy clusters to measure their temperatures and compare them with measurements in identical regions from \chandra\ and \xmm, as defined in \cite{Schellenberger}.
These results allowed an investigation of the relative cross-calibration between \nustar, \chandra, and \xmm.
We found discrepant temperatures when compared to either \chandra\ (Figure ~\ref{fig:temp_plot}) or \xmm\ (Figure ~\ref{fig:temp_plot_xmm}), but in different senses and with different trends.
To assess the potential effect from multi-temperature gas, we fit the \nustar\ spectra of each cluster in several narrower bands and found no evidence of systematic bandpass effects in the temperature measurements. 
Even so, the presence of low temperature gas, detectable in \xmm\ spectra but essentially invisible to \nustar's harder bandpass, could explain the systematically lower temperatures measured by \xmm.
The trend in the \chandra--\nustar\ relation, however, can only be explained by a calibration discrepancy and potentially suggests a modification to \chandra's hard band effective area; spectral simulations of such a modification yield a simple exponential cutoff form for the temperature relation, which is fit to our and other cluster measurements (Figure ~\ref{fig:crosscalib}).
We find a $\sim$5\% increase in \chandra's effective area at $\sim$5~keV is sufficient to explain the observed trend, although an assessment of the exact form of the correction is beyond the scope of this work.
Alternatively, we could presume that \nustar's calibration is off.
Keeping its effective area at 10 keV fixed, relative increases of 18\% and 7\% at 3~keV and 5~keV, respectively, are needed to bring \nustar\ temperature measurements into agreement with \chandra's.
Such a substantial adjustment to \nustar's effective area is unjustifiable given the verification of \nustar's area with stray light observations \citep{Madsen2017crab}, which disallow relative changes across the 3--10~keV band by more than a percent or two.



Lingering systematic uncertainties in the absolute temperatures of massive clusters lead to comparable uncertainties in their estimated masses under the assumption of hydrostatic equilibrium.
These uncertain mass estimates limit their competitiveness for making precision measurements of cosmological parameters, potentially biasing them if systematic errors are the culprit.
In this work, we take advantage of \nustar's excellent calibration to address cross-calibration issues between X-ray observatories, shedding new light on a potential origin for the long-standing discrepancy.

\section{Acknowledgements} 

FL, DRW, and CP acknowledge support from the \nustar\ NASA JPL subcontract RSA1688622, NASA grant 80NSSC20K1000, and the NSF-supported REU program at the University of Utah. 
BJM acknowledges support from Science and Technology Facilities Council grants ST/V000454/1 and ST/Y002008/1.
This work made use of data from the \nustar\ mission, which is led by the California Institute of Technology, managed by the Jet Propulsion Laboratory, and funded by the National Aeronautics and Space Administration. In addition, it employs a list of Chandra datasets, obtained by the Chandra X-ray Observatory, contained in the Chandra Data Collection ~\dataset[DOI: 10.25574]{https://doi.org/10.25574/cdc.418}.




\bibliography{references}
\bibliographystyle{aasjournal}




\end{document}